\documentclass[12pt]{article}

\usepackage[T1]{fontenc}
\usepackage{times}
\usepackage{microtype}
\usepackage[a4paper,top=1in,bottom=1in,left=0.8in,right=0.8in]{geometry}
\usepackage{setspace}
\usepackage{amsmath,amssymb,amsthm}
\usepackage[authoryear,round]{natbib}
\usepackage{authblk}

\usepackage{graphicx}
\usepackage{xcolor}
\providecommand{\widebar}[1]{\overline{#1}}

\theoremstyle{plain}
\newtheorem{theorem}{Theorem}
\newtheorem{proposition}{Proposition}
\theoremstyle{remark}
\newtheorem{remark}{Remark}

\DeclareMathOperator{\pr}{pr}

\newcommand{\expit}{\text{expit}}
\newcommand{\eff}{\text{eff}}
\newcommand{\DR}{\text{DR}}

\newcommand{\sop}{\text{sop}}
\newcommand{\ps}{\text{ps}}
\newcommand{\union}{\text{union}}
\newcommand{\SR}{\text{SR}}
\newcommand{\SOP}{\text{SOP}}
\newcommand{\SSR}{\text{SSR}}
\newcommand{\CIR}{\text{CIR}}
\newcommand{\OP}{\text{OP}}
\newcommand{\sr}{\text{sr}}
\newcommand{\cir}{\text{cir}}
\newcommand{\SCIR}{\text{SCIR}}
\newcommand{\cen}{\text{cen}}

\newcommand{\ind}{\rotatebox[origin=c]{90}{$\models$}}
\usepackage{booktabs}
\usepackage{threeparttable}
\usepackage{placeins}
\usepackage[hidelinks]{hyperref}
\providecommand{\theHtheorem}{\arabic{theorem}}
\providecommand{\theHproposition}{\arabic{proposition}}
\providecommand{\theHremark}{\arabic{remark}}
\usepackage{multirow}
\usepackage{subcaption}

\begin{document}

\title{Towards the Revival of Survival Ratio}

\author[1]{Ruixuan Zhao}
\author[2,3,4]{Jingxin Yan}
\author[4]{Xiaochuan Shi}
\author[5]{Torben Martinussen}
\author[6]{\mbox{Eric Tchetgen Tchetgen}}
\author[1,4]{Linbo Wang}
\affil[1]{Department of Computer and Mathematical Sciences, University of Toronto Scarborough}
\affil[2]{State Key Laboratory of Mathematical Sciences, Academy of Mathematics and Systems Science, \protect\\ Chinese Academy of Sciences}
\affil[3]{School of Mathematical Sciences, University of Chinese Academy of Sciences}
\affil[4]{Department of Statistical Sciences, University of Toronto}
\affil[5]{Section of Biostatistics, Department of Public Health, University of Copenhagen}
\affil[6]{Department of Statistics and Data Science, The Wharton School, University of Pennsylvania}
\date{}

\maketitle

\begin{abstract}
\small
The hazard ratio is the most commonly reported measure in survival analysis, yet it suffers from well-documented limitations, including non-collapsibility and susceptibility to selection bias as treatment-specific risk sets change over time. In this article, we revisit the survival ratio and the cumulative incidence ratio as alternative measures that directly compare survival and failure probabilities. We introduce a novel modeling framework that targets these ratio functions and constructs coherent nuisance models so that the target and nuisance parameters are variation independent. The framework yields valid monotone survival and cumulative distribution functions and supports likelihood-based and doubly robust estimation of the target parameters, together with estimation of covariate-specific survival and cumulative incidence curves. We illustrate the proposed frameworks using two clinical-trial datasets. In a prostate cancer trial, the estimated survival ratio for 1.0 mg estrogen versus placebo increased over follow-up, while a descriptive selected-cohort analysis of a breast cancer screening trial yielded a cumulative incidence ratio that crossed one over time.
\end{abstract}


\section{Introduction}

In survival analysis, the hazard ratio is the most widely used measure to quantify the effect of a binary treatment on a time-to-event outcome. It reflects the ratio of the instantaneous event rates between treatment arms and is typically estimated via the Cox proportional hazards model \citep{cox1972regression}. Hazard ratios may vary over time, making a single hazard-ratio summary difficult to interpret \citep{hernan2010hazards}. Cox regression with time-varying hazard ratios \citep{grambsch1994proportional} can characterize this temporal variation and has been widely used in applied research, such as prominent work on COVID-19 vaccine durability \citep{lin2022effectiveness}. Despite their popularity, time-varying hazard ratios remain difficult to interpret causally. At each time point during follow-up, the hazard ratio compares the treatment arms among individuals who have remained event-free up to that time. The resulting treatment-specific risk sets generally represent different underlying populations. Except under restrictive conditions, comparing them therefore induces survivor-selection bias \citep{hernan2010hazards,martinussen2020subtleties}. Their non-collapsibility further complicates interpretation \citep{greenland1999confounding,martinussen2013collapsibility}. This continued reliance on hazard ratios highlights a methodological gap of practical importance: investigators need models of time-varying survival contrasts that do not condition on treatment-specific risk sets.

In light of these limitations, measures based on survival curves provide alternative ways to assess treatment effects on survival outcomes. Such measures include the survival  difference, the survival  ratio, and the difference in restricted mean survival time. Comparisons on the time scale can also be made using the survival time ratio, which can be estimated using accelerated failure time models (\citealp[Section~5.2]{cox1984analysis};
\citealp{robins1992semiparametric}).

In this article, we focus on the survival ratio, defined as the ratio of survival probabilities between two treatment arms. Survival ratios are attractive from a causal perspective: when defined in terms of potential survival probabilities, they compare the same target population across treatment arms and may have a causal interpretation under appropriate identification conditions. Our contribution, however, is statistical: we develop models and estimators for ratio functions induced by observed conditional survival distributions. Formulating an appropriate model that allows these ratios to vary with covariates is important in many epidemiological and pharmaceutical studies \citep{hingorani2013prognosis,kent2020predictive,rothwell2005subgroup}. The additive hazards framework can induce a parametric model for the survival ratio \citep{aalen1989linear,martinussen2006dynamic}. However, unconstrained estimation may yield negative fitted hazards, causing diagnostics based on martingale residuals to suggest poor fit even when the covariate terms are correctly specified. The monotone modifications proposed by \citet{lin1994semiparametric} can restore valid baseline cumulative hazards and survival functions, but applying them separately by treatment arm need not preserve the modeled survival-ratio relationship.  More fundamentally, the nonnegativity constraints on the two treatment-specific hazards impose joint restrictions on the target parameters indexing the survival ratio and the nuisance parameters indexing the remaining hazard structure, so the two parameter sets are not variation independent. \citet{dukes2019doubly} further developed doubly robust estimators under a class of semiparametric additive hazards models, but this does not resolve the lack of variation independence.

To formalize the modeling problem, let \(A\in\{0,1\}\) denote a binary treatment, \(X\) a vector of covariates, and \(T\in(0,+\infty)\) a survival time. Let \(S_a(t;X)=\pr(T>t\mid A=a,X)\), \(a\in\{0,1\}\), denote the conditional survival function in treatment group \(a\). At each time \(t\), define the conditional survival ratio as
\[
\SR(t;X):=\frac{S_1(t;X)}{S_0(t;X)}.
\]
Modeling \(\SR(t;X)\) over follow-up therefore amounts to modeling a sequence of time-specific relative risks of survival. For binary outcomes, \citet{richardson2017modeling} paired the relative risk with the log odds product as a nuisance parameter, obtaining a variation-independent parameterization of the two outcome probabilities. Applying the same idea at a given time \(t\), define
\(
\OP(t;X):=\frac{S_1(t;X)S_0(t;X)}{(1-S_1(t;X))(1-S_0(t;X))},
\)
so that \((\SR(t;X),\OP(t;X))\) provides a variation-independent parameterization of  survival probabilities \((S_1(t;X),S_0(t;X))\).
However, extending this construction from a single time point to the entire follow-up period is challenging because each underlying survival function must satisfy the following additional restrictions: for \(a\in\{0,1\}\),
\begin{align}
    &S_a(0;X)=1,\qquad \lim_{t\to+\infty}S_a(t;X)=0,\label{Sfun-boundary-3}\\
    &S_a(t;X)\text{ is non-increasing in }t. \label{Sfun-nonin-3}
\end{align}

To address this challenge, we propose a novel modeling framework that directly targets the survival ratio function and develops a coherent set of nuisance models to ensure valid monotone survival functions while preserving variation independence between the target and nuisance parameters. We further develop doubly robust estimators of the parameters indexing the proposed survival-ratio model. When the event of interest is rare during the study period, the survival ratio may remain close to one despite substantial relative differences in cumulative incidence. We therefore introduce an alternative framework that models the cumulative incidence ratio over time, along with the corresponding nuisance models and estimation procedures.

The rest of the article is organized as follows. Sections~\ref{sec:SR} and \ref{sec:est} introduce the survival-ratio modeling framework and develop likelihood-based and doubly robust estimation, respectively. The extension to cumulative incidence ratios is presented in Section~\ref{sec:cir}, after which Section~\ref{sec:sim} assesses the finite-sample performance of estimators for survival ratio and cumulative incidence ratio. Section~\ref{sec:data} illustrates the two frameworks using clinical trial data. We end with a brief discussion in Section~\ref{sec:discussion}.

\section{Survival ratio modeling}\label{sec:SR}

Consider the following survival ratio model
\begin{align*}
\mathcal{M}_{\sr}:\SR(t;X)= \exp(\Phi(t;X,\alpha)),
\end{align*}
where $\Phi(t;X,\alpha)$ is a function of $X$ and $t\in[0,+\infty)$ indexed by an unknown $p_{\alpha}$-dimensional parameter of interest $\alpha\in\mathbb{R}^{p_\alpha}$. Because the survival ratio is induced by the conditional survival functions, $\mathcal{M}_{\sr}$ must be compatible with the conditions in \eqref{Sfun-boundary-3}--\eqref{Sfun-nonin-3}. In particular, \eqref{Sfun-boundary-3} gives $\SR(0;X) =1$, which implies $\Phi(0; X, \alpha)=0$ in $\mathcal{M}_{\sr}$. We provide some example choices of $\Phi(t;X,\alpha)$ below.
 \begin{remark}
(Choices of $\Phi(t; X, \alpha)$)
We consider $\Phi(t; X, \alpha)=\alpha^\top f_{\sr}(t;X)$ in this article, where $f_{\sr}(t;X)$ is a known function of $X$ and $t$, and provide the following examples of $f_{\sr}(t;X)$:
\begin{itemize}
\item[(a).] $f_{\sr}(t;X) = t\cdot X$, which implies $\SR(t;X) \rightarrow + \infty$ if $\alpha^\top X>0$, and $\SR(t;X) \rightarrow 0$ if $\alpha^\top X<0$, as $t\rightarrow + \infty$. In addition, we can expand $f_{\sr}(t;X)$ to include higher-order terms in $t$, such as $f_{\sr}(t;X) = (t \cdot X^\top, t^2 \cdot X^\top)^\top$.
\item[(b).] $f_{\sr}(t;X) = \exp(-1/t) \cdot  X $, which implies $\SR(t;X) \rightarrow \exp(\alpha^\top X)$, as $t\rightarrow +\infty$.
\end{itemize}
The choice of $\Phi(t;X,\alpha)$ determines how the survival ratio evolves over time. Under example (a), the survival ratio tends to zero or infinity, indicating an increasing survival advantage for one treatment over long-term follow-up; under example (b), it tends to a finite constant, indicating a stable relative survival contrast.
\end{remark}

To ensure compatibility with the monotonicity constraint \eqref{Sfun-nonin-3}, we work on a prespecified time grid fixed independently of the observed event times. Specifically, for the finite time points $0=t_0 < t_1 < t_2 < \ldots < t_m$, \eqref{Sfun-nonin-3} implies that $\frac{S_a(t_k;X)}{S_a(t_{k-1};X)}\in[0,1]$ for any $k=1,\ldots,m$ and $a=0,1$. Model $\mathcal{M}_{\sr}$ implies the following sequential target model
\begin{align}\label{eq:SRRX-3}
\SSR(t_{k},t_{k-1}; X) :=
\frac{\SR(t_k;X)}{\SR(t_{k-1};X)}
= \exp(\Psi(t_{k},t_{k-1} ; X, \alpha)),
\end{align}
where $\Psi(t_{k},t_{k-1} ; X, \alpha) = \Phi(t_k; X, \alpha)-\Phi(t_{k-1}; X, \alpha)$. We aim to construct a sequential nuisance model that is variation independent of $\log \SSR(t_k,t_{k-1}; X)$. Motivated by \cite{richardson2017modeling}, consider the following sequential odds-product nuisance model
\begin{align*}
\mathcal{M}_{\sop}:\SOP(t_{k},t_{k-1};X) &:= \frac{\frac{S_1(t_{k}; X)}{S_1(t_{k-1}; X)} \frac{S_0(t_{k}; X)}{S_0(t_{k-1}; X)}}{\Big (1- \frac{S_1(t_{k}; X)}{S_1(t_{k-1}; X)}\Big )\Big (1- \frac{S_0(t_{k}; X)}{S_0(t_{k-1}; X)}\Big )} = \exp(B(t_{k}, t_{k-1}; X, \beta)),
\end{align*}
where $B(t_{k}, t_{k-1};X,\beta)$ is a function of $X$, $t_k$ and $t_{k-1}$ with the unknown $p_{\beta}$-dimensional nuisance parameter $\beta\in \mathbb{R}^{p_{\beta}}$.

\begin{remark}\label{remark:B}
(Choice of $B(t_{k},t_{k-1};X,\beta)$) The conditions in \eqref{Sfun-boundary-3}--\eqref{Sfun-nonin-3}, together with the right-continuity of the survival functions, imply the following limiting properties of $\SOP(t_{k},t_{k-1};X)$. For fixed $t_{k-1}$ such that \(S_a(t_{k-1};X)>0\), \(a=0,1\),
\begin{align*}
	\SOP(t_k,t_{k-1};X) &\rightarrow 0,\quad
	\text{as }t_k\rightarrow+\infty,\\
	\SOP(t_k,t_{k-1};X) &\rightarrow+\infty,\quad
	\text{as }t_k\rightarrow t_{k-1}^{+}.
\end{align*}
Motivated by these properties, we choose $\SOP(t_{k},t_{k-1};X)=\exp(\beta^\top X)/(t_{k} - t_{k-1})$, corresponding to $B(t_{k},t_{k-1};X,\beta)=\beta^\top X - \log(t_{k} - t_{k-1})$.
\end{remark}

Combining the sequential target model in \eqref{eq:SRRX-3} and $\mathcal{M}_{\sop}$, we obtain the following smooth bijection map from $\mathbb{R}^2$ to $(0,1)\times (0,1)$
\begin{align*}
\Big (\Psi(t_{k}, t_{k-1}; X, \alpha), B(t_{k},t_{k-1}; X, \beta) \Big ) \longrightarrow \Big( \frac{S_0(t_{k};X)}{S_0(t_{k-1} ; X)}, \frac{S_1(t_{k};X)}{S_1(t_{k-1} ; X)} \Big ).
\end{align*}
For $k=1,\cdots, m$, we solve the quadratic equations to derive the inverse map and finally obtain
\begin{align}\label{eq:S0}
S_a(t_{k}; X) = \exp\left[\sum_{j=1}^{k} \Big\{ \Gamma(\Psi(t_{j} , t_{j-1}; X, \alpha), B(t_{j},t_{j-1}; X, \beta))  + a\Psi(t_{j} , t_{j-1}; X, \alpha) \Big\} \right ],
\end{align}
where, for $b\neq0$,
\begin{align*}
\Gamma(\psi,b)
&=\log\left[\frac{-(\exp(\psi)+1)\exp(b)+\sqrt{D(\psi,b)}}{2\exp(\psi)(1-\exp(b))}\right],\\
D(\psi,b)
&=\exp(2b)(\exp(\psi)+1)^2+4\exp(\psi+b)(1-\exp(b)).
\end{align*}
For $b=0$, $\Gamma(\psi,0)=-\log\{1+\exp(\psi)\}$, and $a=0,1$.

\section{Estimation}\label{sec:est}

In the analysis of time-to-event data, the survival time $T$ is often right-censored by a censoring time $C$. We first consider survival data recorded on a discrete time grid $0=t_0<t_1<\cdots<t_m$, as commonly arises when event and censoring information is collected at scheduled follow-up visits. Let $Y=\min(T,C)\in\{t_1,\ldots,t_m\}$ denote the recorded time and $\Delta=\mathbb{I}(T\leq C)$ the event indicator. The value $C=t_m$ represents administrative censoring, and the underlying event time $T$ may exceed $t_m$ for individuals who remain event-free at the end of follow-up. For continuously recorded data, we fix the grid a priori, independently of the observed event times, map the observed time $Y$ to this grid, and retain the original event indicator $\Delta$. We then use the discrete-time likelihood below as a grouped-data working model. In what follows, $C$ denotes the latent grid-valued censoring time in the discrete-time data-generating model or working representation. The observed data contain $Y$ and $\Delta$ rather than $C$ itself: when $\Delta=0$, the recorded time $Y$ is the grouped censoring time, whereas when $\Delta=1$, $C$ is known only to satisfy $C\geq Y$. Assume that $C \ind (A,T) \mid X$ and let the conditional probability mass function of $C$ given $X$ be $g(t;X)=\pr(C=t \mid X)$. Let $G(t_k;X)= \sum_{i=1}^{k-1}g(t_i;X)$.

Under $\mathcal{M}_{\sr} \cap \mathcal{M}_{\sop}$, the parameters $\alpha$ and $\beta$ can be estimated by maximum likelihood. The log-likelihood for one observation on the discrete time grid is given by
\begin{align*}
\ell_{\sr}(\alpha,\beta)=\sum_{k=1}^{m}\Big[&\mathbb{I}(Y=t_k)\Delta
\log\big\{[S_A(t_{k-1};X)-S_A(t_k;X)] [1-G(t_k;X)]\big\}\\
&+\mathbb{I}(Y=t_k)(1-\Delta)
\log\{S_A(t_k;X)g(t_k;X)\}\Big].
\end{align*}
We denote the corresponding estimators by $\widehat{\alpha}$ and $\widehat{\beta}$. The score equations and sandwich variance estimator are given in \ref{supp-sec:sr-mle} of the Supplementary Materials. Under standard regularity conditions \citep{white1982maximum}, \(\widehat{\alpha}\) converges to a pseudo-true limit \(\alpha^*\), which equals the true value \(\dot{\alpha}\) when both models are correctly specified, but may differ from \(\dot{\alpha}\) if \(\mathcal{M}_{\sop}\) is misspecified. To overcome this potential inconsistency, we develop a robust estimation method based on semiparametric theory \citep{bickel1993efficient,tsiatis2006semiparametric}.

For implementation, we use the logistic propensity score working model
\begin{align*}
\mathcal{M}_{\ps}:e(X):=\pr(A=1 \mid X) = e(X;\gamma),
\end{align*}
where $e(X;\gamma)=\expit (\gamma^\top X)$ and $\gamma$ is an unknown $p_{\gamma}$-dimensional parameter. This logistic link is a working-model choice, not a requirement of the efficient-score construction. Let $\widehat{\gamma}$ denote its maximum likelihood estimator, which converges in probability to a pseudo-true limit $\gamma^*$ that equals the true value $\dot\gamma$ when $\mathcal{M}_{\ps}$ holds.
We assume the positivity condition that with probability $1$, for some fixed, sufficiently small $\epsilon>0$, $e(X), \frac{S_a(t_k;X)}{S_a(t_{k-1};X)}\in (\epsilon,1-\epsilon)$ for all $a=0,1$ and $k=1,...,m$. The following theorem characterizes the efficient score for \(\alpha\) and provides a strategy for constructing a robust estimator.

\begin{theorem}\label{thm:sur:effscore}
Under the observed-data law restricted by the model $\mathcal{M}_{\sr}$, the efficient score for $\alpha$ is
\begin{align*}
S^*(\alpha) = \mathcal{W}_{\text{eff}}(X) R(\alpha)(A-e(X)),
\end{align*}
where $\mathcal{W}_{\eff}(X)$ is the efficient $p_\alpha\times m$ weighting matrix, and $R(\alpha) = (R_1(\alpha),...,R_m(\alpha))^\top$ with
\begin{align*}
R_k(\alpha) &= \frac{1}{S_0(t_k;X)} H_k(\alpha) + \frac{1}{S_0(t_k;X)} C_k(\alpha)
- \frac{1}{S_0(t_{k-1};X)} H_{k-1}(\alpha),\\
H_k(\alpha) &= \exp(-\Phi(t_k;X,\alpha) A) \prod_{j=0}^{k} (1-\mathbb{I}(Y=t_j))
\end{align*}
and
\[
C_k(\alpha) = \exp(-\Phi(t_k;X,\alpha)A)(1-\Delta)\mathbb{I}(Y=t_k).
\]
\end{theorem}

\begin{remark}\label{remark:sr-censoring}(Censoring and the efficient score)
Under $C\ind(A,T)\mid X$, the probability of remaining uncensored at $t_k$, $1-G(t_k;X)$, is the same for both treatment arms. In deriving
$S^*(\alpha)$, this probability multiplies both treatment-specific terms and cancels from their contrast, so $S^*(\alpha)$ does not
involve the censoring distribution $G(t;X)$. If censoring depends on $A$ conditional on $X$, this cancellation need not occur and the efficient score
generally involves the censoring distribution; see Proposition \ref{supp-prop:effScore} in \ref{supp-sec:sr-censoring} of the Supplementary Materials.
\end{remark}

By Theorem \ref{thm:sur:effscore}, estimating $\alpha$ by directly solving the estimating equation induced by $S^*(\alpha)$ without any parametric modeling is generally infeasible, particularly when $X$ is high-dimensional and continuous. For unknown components $S_0(t_k;X)$ and $e(X)$ appearing in $S^*(\alpha)$, we specify $\mathcal{M}_{\sr}\cap \mathcal{M}_{\sop}$ to derive $S_0(t_k;X,\alpha,\beta)$ from \eqref{eq:S0}, and, under $\mathcal{M}_{\ps}$, we parameterize $e(X)$ as $e(X;\gamma)$. Then, we construct the estimating equation by plugging \(S_0(t_k;X,\widehat{\alpha},\widehat{\beta})\) and \(e(X;\widehat{\gamma})\) into \(S^\ast(\alpha)\), while treating the \(\alpha\) appearing in \(H_k(\alpha)\) and \(C_k(\alpha)\) as the free argument of the estimating equation. Specifically, the doubly robust estimator $\widehat{\alpha}_{\DR}(\mathcal{W})$ is obtained by solving
\begin{align}\label{eq:effscore}
0=\mathbb{P}_n [U_{\sr}(\alpha;\widehat{\alpha},\widehat{\beta},\widehat{\gamma};\mathcal{W})],
\end{align}
where $\mathbb{P}_n$ denotes the empirical average and
\[
U_{\sr}(\alpha;\widehat{\alpha},\widehat{\beta},\widehat{\gamma};\mathcal{W}) = \mathcal{W}(X)R(\alpha;\widehat{\alpha},\widehat{\beta})(A-e(X;\widehat{\gamma}))
\]
with $p_\alpha\times m$ weighting matrix $\mathcal{W}(X)$,
and $R(\alpha;\widehat{\alpha},\widehat{\beta})=(R_1(\alpha;\widehat{\alpha},\widehat{\beta}),\cdots,R_m(\alpha;\widehat{\alpha},\widehat{\beta}))^\top$ with $R_k(\alpha;\widehat{\alpha},\widehat{\beta})=\frac{1}{S_0(t_k;X,\widehat{\alpha},\widehat{\beta})} H_k(\alpha) + \frac{1}{S_0(t_k;X,\widehat{\alpha},\widehat{\beta})} C_k(\alpha) - \frac{1}{S_0(t_{k-1};X,\widehat{\alpha},\widehat{\beta})} H_{k-1}(\alpha)$. We restrict attention to choices of $\mathcal{W}(X)$ for which the Jacobian of the expected estimating function with respect to $\alpha$ is nonsingular at $\dot\alpha$. As the efficient weighting matrix $\mathcal{W}_{\eff}(X)$ depends on $e(X;\gamma)$ and $S_0(t_k;X,\alpha,\beta)$, we obtain the estimator $\widehat{\alpha}_{\DR}(\mathcal{W}_{\eff})$ by solving \eqref{eq:effscore} with the plug-in efficient weight $\mathcal{W}(X)=\mathcal{W}_{\eff}(X;\widehat{\alpha},\widehat{\beta},\widehat{\gamma})$. Proposition \ref{prop:sur:dr} summarizes the properties of $\widehat{\alpha}_{\DR}(\mathcal{W})$.

\begin{proposition}\label{prop:sur:dr}
Under standard regularity conditions and the positivity condition, $\widehat{\alpha}_{\DR}(\mathcal{W})$ is a consistent and asymptotically normal (CAN) estimator of $\alpha$ in the union model $\mathcal{M}_{\union}:= \mathcal{M}_{\sr}\cap \{\mathcal{M}_{\sop} \cup \mathcal{M}_{\ps}\}$.
Furthermore, if both $\mathcal{M}_{\sop}$ and $\mathcal{M}_{\ps}$ are correctly specified,  $\widehat{\alpha}_{\DR}(\mathcal{W}_{\eff})$ attains the semiparametric efficiency bound under $\mathcal{M}_{\union}$.
\end{proposition}

The explicit form of $\mathcal{W}_{\eff}(X)$, the construction of the plug-in estimator $\widehat{\alpha}_{\DR}(\mathcal W_{\eff})$, and the influence-function-based variance estimator are given in \ref{supp-sec:sr-dr} of the Supplementary Materials.

\section{Cumulative incidence ratio modeling}\label{sec:cir}

When the event of interest is rare, survival probabilities under both treatment arms remain close to one, so \(\SR(t;X)\) may obscure substantial relative differences in cumulative incidence. The cumulative incidence ratio directly compares the risks of experiencing the event by time \(t\) and is therefore more informative in such settings. Extending our framework to the cumulative incidence ratio requires more than replacing survival functions by cumulative distribution functions. First, because a cumulative distribution function is anchored at one as $t\rightarrow +\infty$ rather than at the beginning of follow-up, we adopt a backward sequential construction. Second, unlike for the survival ratio under \(C\ind(A,T)\mid X\), the censoring distribution enters the efficient score and changes the conditions for robust estimation.

We consider the following cumulative incidence ratio model
\begin{align*}
\mathcal{M}_{\cir}:\CIR(t;X) := \frac{F_1(t;X)}{F_0(t;X)} = \exp(\Lambda(t;X,\rho)),
\end{align*}
where $F_a(t;X)=\pr(T\leq t \mid A=a, X)=1-S_a(t;X)$ is the conditional cumulative distribution function of $T$ under $A=a$, and $\rho\in\mathbb{R}^{p_\rho}$ is the finite-dimensional parameter of interest. Here and throughout, cumulative incidence refers to this single-event distribution function, rather than to a cause-specific cumulative incidence function in a competing-risks setting. We use $\Lambda(t;X,\rho)=\rho^\top f_{\cir}(t;X)$ for a known function $f_{\cir}(t;X)$.

Each $F_a(t;X)$ starts at zero, converges to one, and is non-decreasing in $t$. Consequently, the forward construction used for the survival ratio cannot be initialized at $t=0$, where $F_a(0;X)=0$ for $a\in\{0,1\}$. We instead add a terminal point $t_{m+1}\rightarrow+\infty$, set $F_a(t_{m+1};X)=1$, and work backward over $0<t_1<\cdots<t_m<t_{m+1}$. Let $q_{a,k}(X)=F_a(t_k;X)/F_a(t_{k+1};X)$ for $k=1,...,m$. The target model implies
\begin{align}\label{model:SCIR}
\SCIR(t_{k+1},t_k;X):=\frac{\CIR(t_k;X)}{\CIR(t_{k+1};X)} =\frac{q_{1,k}(X)}{q_{0,k}(X)}= \exp\{\Upsilon(t_{k+1},t_k;X,\rho)\},
\end{align}
where $\Upsilon(t_{k+1},t_k;X,\rho)=\Lambda(t_k;X,\rho)-\Lambda(t_{k+1};X,\rho)$. We pair this target with the backward odds-product nuisance model
\begin{align*}
\widetilde{\mathcal{M}}_{\sop}:\quad
\widetilde{\SOP}(t_{k+1},t_k;X)
:=\frac{q_{1,k}(X)q_{0,k}(X)}{\{1-q_{1,k}(X)\}\{1-q_{0,k}(X)\}}
=\exp\{\Xi(t_{k+1},t_k;X,\tau)\},
\end{align*}
where $\tau\in \mathbb{R}^{p_{\tau}}$ is a finite-dimensional nuisance parameter. For each $k$, \eqref{model:SCIR} and $\widetilde{\mathcal{M}}_{\sop}$ give a smooth bijection map
\begin{align*}
\big(\Upsilon(t_{k+1},t_k;X,\rho),\Xi(t_{k+1},t_k;X,\tau)\big)
\longrightarrow \big(q_{0,k}(X),q_{1,k}(X)\big)\in(0,1)^2.
\end{align*}
Starting from $F_a(t_{m+1};X)=1$ and applying this map recursively yields
\[
F_a(t_k;X)=\exp\left[\sum_{j=k}^m\left\{\Gamma(\Upsilon(t_{j+1},t_j;X,\rho),\Xi(t_{j+1},t_j;X,\tau))+a\Upsilon(t_{j+1},t_j;X,\rho)\right\}\right].
\]
Thus, the backward construction preserves the terminal boundary and monotonicity of the cumulative distribution functions. Valid choices, including
\(
f_{\cir}(t;X)=X/t
\)
and
\(
\Xi(t_{k+1},t_k;X,\tau)=\tau^\top X-\log(1/t_k-1/t_{k+1}),
\)
are discussed in Remark \ref{supp-remark:cir} in \ref{supp-sec:cir-model} of the Supplementary Materials.

Under $\mathcal{M}_{\cir}\cap\widetilde{\mathcal{M}}_{\sop}$, the parameters $\rho$ and $\tau$ can be estimated by maximum likelihood. The log-likelihood for one observation on the discrete time grid is given by
\begin{align*}
\ell_{\cir}(\rho,\tau)=\sum_{k=1}^{m}\Big[&\mathbb{I}(Y=t_k)\Delta
\log\big\{[F_A(t_k;X)-F_A(t_{k-1};X)] [1-G(t_k;X)]\big\}\\
&+\mathbb{I}(Y=t_k)(1-\Delta)
\log\{[1-F_A(t_k;X)]g(t_k;X)\}\Big].
\end{align*}
The score equations and sandwich variance estimator are given in \ref{supp-sec:cir-mle} of the Supplementary Materials. Robust estimation is more involved than the backward analogue of Section \ref{sec:est}, and its key distinction appears in the efficient score. To proceed, we also impose the following positivity condition: with
probability one, for some fixed, sufficiently small $\epsilon>0$,
$e(X)$ and $q_{a,k}(X)$ lie in $(\epsilon,1-\epsilon)$ for all
$a=0,1$ and $k=1,\ldots,m$, and
$G(t_m;X)<1-\epsilon$.
\begin{theorem}\label{thm:sur:effscore-CIR}
Under the observed-data law restricted by the model $\mathcal{M}_{\cir}$, the efficient score for $\rho$ is
\begin{align*}
S^*(\rho) = \widetilde{\mathcal{W}}_{\eff}(X)\left[\frac{A}{e(X)}Z_1(\rho)-\frac{1-A}{1-e(X)}Z_0(\rho)\right],
\end{align*}
where $\widetilde{\mathcal{W}}_{\eff}(X)$ is an efficient $p_\rho\times m$ weighting matrix and $Z_a(\rho)$ is an arm-specific $m$-dimensional residual. Their explicit forms are given in \ref{supp-sec:cir-dr} of the Supplementary Materials.
\end{theorem}

The efficient score in Theorem \ref{thm:sur:effscore} also admits an arm-specific representation analogous to that in Theorem \ref{thm:sur:effscore-CIR}, $S^*(\alpha)={\mathcal{W}}_{\eff}(X)e(X)\{1-e(X)\}\left[\frac{A}{e(X)}R_1(\alpha)-\frac{1-A}{1-e(X)}R_0(\alpha)\right]$, where the residual $R_a(\alpha)$ denotes $R(\alpha)$ evaluated at $A=a$. Despite this formal similarity, the arm-specific residuals have a key difference. In contrast to $R_a(\alpha)$, for the cumulative incidence ratio, $Z_a(\rho)$ depends on $G(t;X)$ because the censoring distribution determines which failure increments are observed at each step of the backward construction and therefore cannot be factored out as a common term across treatment arms. Censoring must therefore be modeled when robustness is obtained through treatment and censoring adjustment rather than through a correct $\widetilde{\mathcal{M}}_{\sop}$.

Write $\mathcal{M}_{\cen}$ for a model $G(t;X)=G(t;X,\eta)$. Based on $S^*(\rho)$, we define
$\widehat\rho_{\DR}(\mathcal W)$ as a solution to the estimating
equation obtained by replacing $\widetilde{\mathcal W}_{\eff}(X)$
with a weighting matrix $\mathcal W(X)$ and substituting estimates
for $F_0(t_k;X)$, $e(X)$, and $G(t_k;X)$ obtained under
$\mathcal{M}_{\cir}\cap\widetilde{\mathcal{M}}_{\sop}$,
$\mathcal{M}_{\ps}$, and $\mathcal{M}_{\cen}$, respectively. \ref{supp-sec:cir-dr} of the Supplementary Materials gives the censoring model, the explicit estimating equation defining $\widehat\rho_{\DR}(\mathcal W)$, and the corresponding variance estimator.

\begin{proposition}\label{prop:sur:DR-CIR} Under standard regularity conditions and the positivity condition, the doubly robust estimator  $\widehat{\rho}_{\DR}(\mathcal{W})$ is a CAN estimator of $\rho$ in the union model $\widetilde{\mathcal{M}}_{\union}:=\mathcal{M}_{\cir}\cap \{\widetilde{\mathcal{M}}_{\sop}\cup (\mathcal{M}_{\ps} \cap \mathcal{M}_{\cen})\}$.
Furthermore, if $\widetilde{\mathcal{M}}_{\sop}$,
$\mathcal{M}_{\ps}$, and $\mathcal{M}_{\cen}$ are correctly
specified,  $\widehat{\rho}_{\DR}(\widetilde{\mathcal W}_{\eff})$ attains the semiparametric efficiency bound under $\widetilde{\mathcal{M}}_{\union}$.
\end{proposition}

Thus, a correct backward odds-product model is sufficient for consistency; if that model is misspecified, both the propensity-score and censoring models must be correct. The additional requirement that the censoring model be correctly specified does not arise in the corresponding survival-ratio result in Proposition \ref{prop:sur:dr}.

\section{Simulation studies}\label{sec:sim}

Two simulation studies are conducted to evaluate our proposed estimators for the survival ratio and cumulative incidence ratio models, respectively.

\subsection{Survival ratio model}\label{sec:sr-sim}

We evaluate the finite-sample performance of the estimators proposed in Section \ref{sec:est}. The simulated data are generated on a prespecified time grid $(t_1,...,t_m)$ under the models $\mathcal{M}_{\sr}$, $\mathcal{M}_{\sop}$ and $\mathcal{M}_{\ps}$ with true parameter values $\dot\alpha=(0,1)^{\top}$,  $\dot\beta=(-0.5,1)^{\top}$ and $\dot\gamma=(0.1,-0.5)^{\top}$. In particular, we set $f_{\sr}(t;X)=tX$ and $B(t_k,t_{k-1};X,\beta)=\beta^\top X-\log(t_k-t_{k-1})$, where
$X=(1,X_1)^\top$ with $X_1 \sim \mathrm{Unif}(-2,2)$. Under this specification, $\log{\SR(t;X)}=tX_1$, so the survival ratio $\SR(t;X)$ varies with both follow-up time and the covariate $X_1$. We then generate the treatment $A$ according to
\(e(X;\dot\gamma)=\operatorname{expit}(\dot\gamma^\top X).\) The survival time $T$ is then sampled under $\mathcal{M}_{\sr}\cap \mathcal{M}_{\sop}$, and the censoring time $C$ is generated from an exponential distribution with rate \(2/t_m\) and discretized on the prespecified time grid, with administrative censoring at \(t_m\).

We also consider the scenarios in which either or both of the nuisance models $\mathcal{M}_{\sop}$ and $\mathcal{M}_{\ps}$ are misspecified. For each misspecified nuisance model, we replace $X$ in the corresponding working model with $X^*=(1,X_1^*)^\top$, where $X_1^* \sim \text{Unif}(-2,2)$ and $X_1^*$ is independent of $(X,A,T,C)$. Separate independent copies of \(X^*\) are used for the two nuisance models when both are misspecified. Specifically, we consider four scenarios where the target model \(\mathcal M_{\sr}\) remains correctly specified:
\begin{enumerate}
	\item \textsf{bth}: use $X$ in $\mathcal{M}_{\sop}$ and $\mathcal{M}_{\ps}$;
	\item \textsf{psc}: use $X$ in $\mathcal{M}_{\ps}$, but $X^*$ in $\mathcal{M}_{\sop}$;
	\item \textsf{sop}: use $X$ in $\mathcal{M}_{\sop}$, but $X^*$ in $\mathcal{M}_{\ps}$;
	\item \textsf{bad}: use $X^*$ in $\mathcal{M}_{\sop}$ and $\mathcal{M}_{\ps}$.
\end{enumerate}
We consider sample sizes $n\in\{500,1000\}$ and use a prespecified time grid with \(m=10\) points. In \emph{Example~1} the time grid is set as  $(1,\dots,10)$, whereas in \emph{Example~2} ten time points are sampled uniformly from $(0,10)$, ordered to form the time grid, and held fixed across replications. Each configuration is replicated $500$ times.

We evaluate the following estimators developed in Section \ref{sec:est}:
\begin{enumerate}
	\item \textsf{mle}: the maximum likelihood estimator (MLE) $\widehat{\alpha}$;
	\item \textsf{dr}: the doubly robust estimator (DR) $\widehat{\alpha}_{\DR}(\mathcal{W}_{\eff})$.
\end{enumerate}
Because maximum-likelihood estimation of $\alpha$ depends only on $\mathcal{M}_{\sop}$, its results are the same under \textsf{bth} and \textsf{sop}, and likewise under \textsf{psc} and \textsf{bad}. We therefore label the two distinct MLE cases as $\textsf{mle.bth}$ and $\textsf{mle.bad}$ and the four DR cases as $\textsf{dr.bth}$, $\textsf{dr.psc}$, $\textsf{dr.sop}$, and $\textsf{dr.bad}$. The standard deviations for $\textsf{mle}$ and $\textsf{dr}$ are estimated using the methods described in \ref{supp-sec:sr-mle} and \ref{supp-sec:sr-dr} of
 the Supplementary Materials, respectively.

Table~\ref{tab:dr500} and Table S1 report, for each estimator and scenario, the Monte Carlo bias and its Monte Carlo standard error (SE), the ratio of the average estimated standard deviation to the Monte Carlo standard deviation (SD accuracy), and the empirical $95\%$ confidence interval coverage (Coverage). Across scenarios in which either \(\mathcal M_{\sop}\) or \(\mathcal M_{\ps}\) is correctly specified, the DR estimator shows small bias and empirical coverage generally close to the nominal 95\% level, consistent with its doubly robust property in Proposition \ref{prop:sur:dr}. When \(\mathcal M_{\sop}\) is misspecified, the bias of the MLE persists as the sample size increases, while its empirical coverage is well below 95\%, particularly for \(\alpha_0\), consistent with convergence to a misspecified pseudo-true value. When both \(\mathcal M_{\sop}\) and \(\mathcal M_{\ps}\) are misspecified, the $\textsf{dr.bad}$ case also shows bias and below-nominal coverage, especially for $\alpha_0$, even though SD accuracy remains close to one. The average estimated standard deviations are generally close to the Monte Carlo standard deviations, especially when either \(\mathcal M_{\sop}\) or \(\mathcal M_{\ps}\) is correctly specified. Finally, when both \(\mathcal M_{\sop}\) and \(\mathcal M_{\ps}\) are correctly specified, the fully parametric MLE attains the parametric efficiency bound, which is no larger than the semiparametric efficiency bound attained by the DR estimator. This ordering is consistent with the Monte Carlo standard error reported for $\textsf{mle.bth}$ and $\textsf{dr.bth}$.

\begin{table}[htbp!]
	\centering
	\caption{Monte Carlo results of the proposed MLE and DR estimators.
		The true values for $\alpha_0$ and $\alpha_1$ are 0 and $1$, respectively,
		for both examples. Sample size = 500}
	\label{tab:dr500}
	\begin{threeparttable}
		\begin{tabular}{llrrrr}
			\toprule
			& & \multicolumn{2}{c}{Example 1} & \multicolumn{2}{c}{Example 2} \\
			\cmidrule(lr){3-4} \cmidrule(lr){5-6}
			& & \multicolumn{1}{c}{$\alpha_0$} & \multicolumn{1}{c}{$\alpha_1$}
			& \multicolumn{1}{c}{$\alpha_0$} & \multicolumn{1}{c}{$\alpha_1$} \\
			\midrule
			\multicolumn{6}{l}{\textbf{Bias (SE)}} \\
			& mle.bth & $-0.002\;(0.005)$ & $\phantom{-}0.006\;(0.005)$ & $\phantom{-}0.007\;(0.005)$ & $\phantom{-}0.002\;(0.006)$ \\
			& mle.bad & $\phantom{-}0.257\;(0.005)$ & $\phantom{-}0.076\;(0.005)$ & $\phantom{-}0.113\;(0.005)$ & $\phantom{-}0.116\;(0.006)$ \\
			& dr.bth & $-0.002\;(0.005)$ & $\phantom{-}0.014\;(0.005)$ & $\phantom{-}0.001\;(0.007)$ & $\phantom{-}0.019\;(0.007)$ \\
			& dr.psc & $-0.003\;(0.005)$ & $\phantom{-}0.016\;(0.005)$ & $-0.001\;(0.007)$ & $\phantom{-}0.021\;(0.007)$ \\
			& dr.sop & $\phantom{-}0.001\;(0.005)$ & $\phantom{-}0.017\;(0.005)$ & $\phantom{-}0.006\;(0.007)$ & $\phantom{-}0.027\;(0.008)$ \\
			& dr.bad & $-0.098\;(0.005)$ & $-0.013\;(0.005)$ & $-0.183\;(0.007)$ & $\phantom{-}0.059\;(0.008)$ \\
			\addlinespace
			\multicolumn{6}{l}{\textbf{SD Accuracy\textsuperscript{*}}} \\
			& mle.bth & 0.984 & 0.981 & 1.037 & 0.996 \\
			& mle.bad & 0.965 & 0.928 & 1.014 & 0.950 \\
			& dr.bth & 0.984 & 0.988 & 0.955 & 1.026 \\
			& dr.psc & 0.975 & 0.965 & 0.948 & 1.009 \\
			& dr.sop & 0.980 & 0.986 & 0.961 & 1.013 \\
			& dr.bad & 0.969 & 0.958 & 0.868 & 0.966 \\
			\addlinespace
			\multicolumn{6}{l}{\textbf{Coverage\textsuperscript{**}}} \\
			& mle.bth & 0.944 & 0.938 & 0.962 & 0.948 \\
			& mle.bad & 0.258 & 0.888 & 0.850 & 0.852 \\
			& dr.bth & 0.940 & 0.946 & 0.930 & 0.952 \\
			& dr.psc & 0.946 & 0.942 & 0.936 & 0.942 \\
			& dr.sop & 0.946 & 0.948 & 0.938 & 0.948 \\
			& dr.bad & 0.836 & 0.928 & 0.718 & 0.934 \\
			\bottomrule
		\end{tabular}
		\begin{tablenotes}
			\item[*] SD Accuracy = Average estimated SD / Monte Carlo SD.
			\item[**] Confidence level = 95\%.
		\end{tablenotes}
	\end{threeparttable}
\end{table}

\subsection{Cumulative incidence model}\label{sec:cir-sim}

We then evaluate the finite-sample performance of the proposed estimators of $\rho$ in the cumulative incidence ratio model. The simulation follows the same general design as that used for the estimators of $\alpha$ in Section \ref{sec:sr-sim}, including the covariate distribution, the sample sizes, and the number of Monte Carlo replications. We therefore focus here on the features specific to the setting of the cumulative incidence ratio model.

The simulated data are generated on a prespecified time grid $(t_1,...,t_m)$ under the models $\mathcal{M}_{\cir}$, $\widetilde{\mathcal M}_{\sop}$, $\mathcal{M}_{\ps}$ and $\mathcal{M}_{\cen}$ with true parameter values $\dot\rho=(0.8,1.4)^{\top}$, $\dot\tau=(-1.5,1.2)^{\top}$, $\dot\gamma=(0.1,-0.3)^{\top}$, and $\dot\eta=(-3.2,1.0)^{\top}$. We set $f_{\cir}(t;X)=X/t$ and $\Xi(t_{k+1},t_k;X,\tau)=\tau^\top X - \log({1}/{t_k} - {1}/{t_{k+1}})$. Under this specification, $\log{\CIR(t;X)}=(0.8+1.4X_1)/t$, so the cumulative incidence ratio $\CIR(t;X)$ varies with both follow-up time and the covariate $X_1$. The survival time $T$ is sampled under $\mathcal M_{\cir}\cap\widetilde{\mathcal M}_{\sop}$, while the censoring time $C$ is generated on the prespecified discrete time grid using the conditional censoring probability
\(
\Pr(C=t_k \mid C \geq t_k, X)
=
\operatorname{expit}(-3.2 + X_1)
\) for $k=1,...,m-1$.

We use the same misspecification mechanism as in Section \ref{sec:sr-sim}. Specifically, whenever a nuisance model is misspecified, $X$ in that model is replaced by $X^*$. When more than one nuisance model is misspecified, separate independent copies of $X^*$ are used for the corresponding models. The target model $\mathcal M_{\cir}$ remains correctly specified throughout, while the nuisance models are considered under the following four configurations:
\begin{enumerate}
	\item \textsf{all}: use $X$ in $\widetilde{\mathcal M}_{\sop}$, $\mathcal{M}_{\ps}$, and $\mathcal{M}_{\cen}$;
	\item \textsf{psc}: use $X$ in $\mathcal{M}_{\ps}$ and $\mathcal{M}_{\cen}$, but $X^*$ in $\widetilde{\mathcal M}_{\sop}$;
	\item \textsf{sop}: use $X$ in $\widetilde{\mathcal M}_{\sop}$, but $X^*$ in $\mathcal{M}_{\ps}$ and $\mathcal{M}_{\cen}$;
	\item \textsf{bad}: use $X^*$ in $\widetilde{\mathcal M}_{\sop}$, $\mathcal{M}_{\ps}$, and $\mathcal{M}_{\cen}$.
\end{enumerate}
We use the time grid $(1,...,10)$ in \emph{Example~1}, but in \emph{Example~2}, ten time points are sampled once from $\mathrm{Unif}(1,10)$ and held fixed across Monte Carlo replications, because the specification $f_{\cir}(t;X)=X/t$ requires the time points to be bounded away from zero. The estimators compared and the performance measures used for evaluation are the same as those in Section \ref{sec:sr-sim}.

The maximum likelihood estimator of $\rho$ depends only on $\widetilde{\mathcal M}_{\sop}$. We therefore report two distinct MLE cases, \textsf{mle.all} and \textsf{mle.bad}, corresponding to correctly specified and misspecified $\widetilde{\mathcal M}_{\sop}$, respectively, and four DR cases, \textsf{dr.all}, \textsf{dr.psc}, \textsf{dr.sop}, and \textsf{dr.bad}. The standard deviations are estimated using the methods described in \ref{supp-sec:cir-mle} and \ref{supp-sec:cir-dr} of the Supplementary Materials.

Table~\ref{tab:cir_500} and Table S2 report the Monte Carlo results for the estimators. The overall patterns are similar to those observed in Section \ref{sec:sr-sim}. Across configurations in which either $\widetilde{\mathcal M}_{\sop}$ is correctly specified or both $\mathcal M_{\ps}$ and $\mathcal M_{\cen}$ are correctly specified, the DR estimator exhibits some finite-sample bias at $n=500$, but the bias decreases substantially as the sample size increases to $n=1000$. At $n=1000$, the biases are generally small and the empirical coverage probabilities are close to the nominal 95\% level. Misspecification of $\widetilde{\mathcal M}_{\sop}$ results in persistent bias and below-nominal coverage for $\rho_1$ for \textsf{mle.bad}. When all nuisance models are misspecified, \textsf{dr.bad} exhibits substantial bias, with particularly pronounced undercoverage for $\rho_0$, even though SD accuracy generally remains close to one. Furthermore, the results are similar under the regular and irregular time grids considered in \emph{Examples~1 and 2}, suggesting that the robustness patterns are not sensitive to grid spacing.

\begin{table}[htbp!]
	\centering
	\caption{Monte Carlo results of the proposed MLE and DR estimators.
		The true values for $\rho_0$ and $\rho_1$ are 0.8 and 1.4, respectively,
		for both examples. Sample size = 500}
	\label{tab:cir_500}
	\begin{threeparttable}
		\begin{tabular}{llrrrr}
			\toprule
			& & \multicolumn{2}{c}{Example 1} & \multicolumn{2}{c}{Example 2} \\
			\cmidrule(lr){3-4} \cmidrule(lr){5-6}
			& & \multicolumn{1}{c}{$\rho_0$} & \multicolumn{1}{c}{$\rho_1$}
			& \multicolumn{1}{c}{$\rho_0$} & \multicolumn{1}{c}{$\rho_1$} \\
			\midrule
			\multicolumn{6}{l}{\textbf{Bias (SE)}} \\
			& mle.all & $-0.022\;(0.032)$ & $\phantom{-}0.096\;(0.026)$ & $-0.030\;(0.033)$ & $\phantom{-}0.064\;(0.025)$ \\
			& mle.bad & $-0.077\;(0.028)$ & $\phantom{-}1.232\;(0.025)$ & $\phantom{-}0.006\;(0.029)$ & $\phantom{-}1.226\;(0.027)$ \\
			& dr.all  & $\phantom{-}0.072\;(0.035)$ & $\phantom{-}0.015\;(0.026)$ & $\phantom{-}0.007\;(0.034)$ & $\phantom{-}0.046\;(0.025)$ \\
			& dr.psc  & $-0.034\;(0.032)$ & $\phantom{-}0.130\;(0.028)$ & $-0.020\;(0.034)$ & $\phantom{-}0.076\;(0.029)$ \\
			& dr.sop  & $\phantom{-}0.061\;(0.034)$ & $\phantom{-}0.010\;(0.026)$ & $\phantom{-}0.007\;(0.034)$ & $\phantom{-}0.039\;(0.026)$ \\
			& dr.bad  & $-0.995\;(0.032)$ & $\phantom{-}0.638\;(0.028)$ & $-1.031\;(0.032)$ & $\phantom{-}0.600\;(0.028)$ \\
			\addlinespace
			\multicolumn{6}{l}{\textbf{SD Accuracy\textsuperscript{*}}} \\
			& mle.all & 0.957 & 0.965 & 0.922 & 0.967 \\
			& mle.bad & 0.961 & 0.972 & 0.944 & 0.968 \\
			& dr.all  & 0.922 & 0.980 & 0.912 & 0.988 \\
			& dr.psc  & 1.032 & 1.031 & 0.960 & 0.963 \\
			& dr.sop  & 0.925 & 0.981 & 0.913 & 0.993 \\
			& dr.bad  & 0.944 & 0.965 & 0.939 & 0.946 \\
			\addlinespace
			\multicolumn{6}{l}{\textbf{Coverage\textsuperscript{**}}} \\
			& mle.all & 0.950 & 0.932 & 0.932 & 0.940 \\
			& mle.bad & 0.936 & 0.392 & 0.930 & 0.442 \\
			& dr.all  & 0.940 & 0.944 & 0.930 & 0.948 \\
			& dr.psc  & 0.962 & 0.948 & 0.940 & 0.936 \\
			& dr.sop  & 0.932 & 0.940 & 0.930 & 0.938 \\
			& dr.bad  & 0.662 & 0.810 & 0.650 & 0.816 \\
			\bottomrule
		\end{tabular}
		\begin{tablenotes}
			\item[*] SD Accuracy = Average estimated SD / Monte Carlo SD.
			\item[**] Confidence level = 95\%.
		\end{tablenotes}
	\end{threeparttable}
\end{table}

\begin{figure}[!htbp]
	\centering
	\includegraphics[width=0.98\linewidth]
	{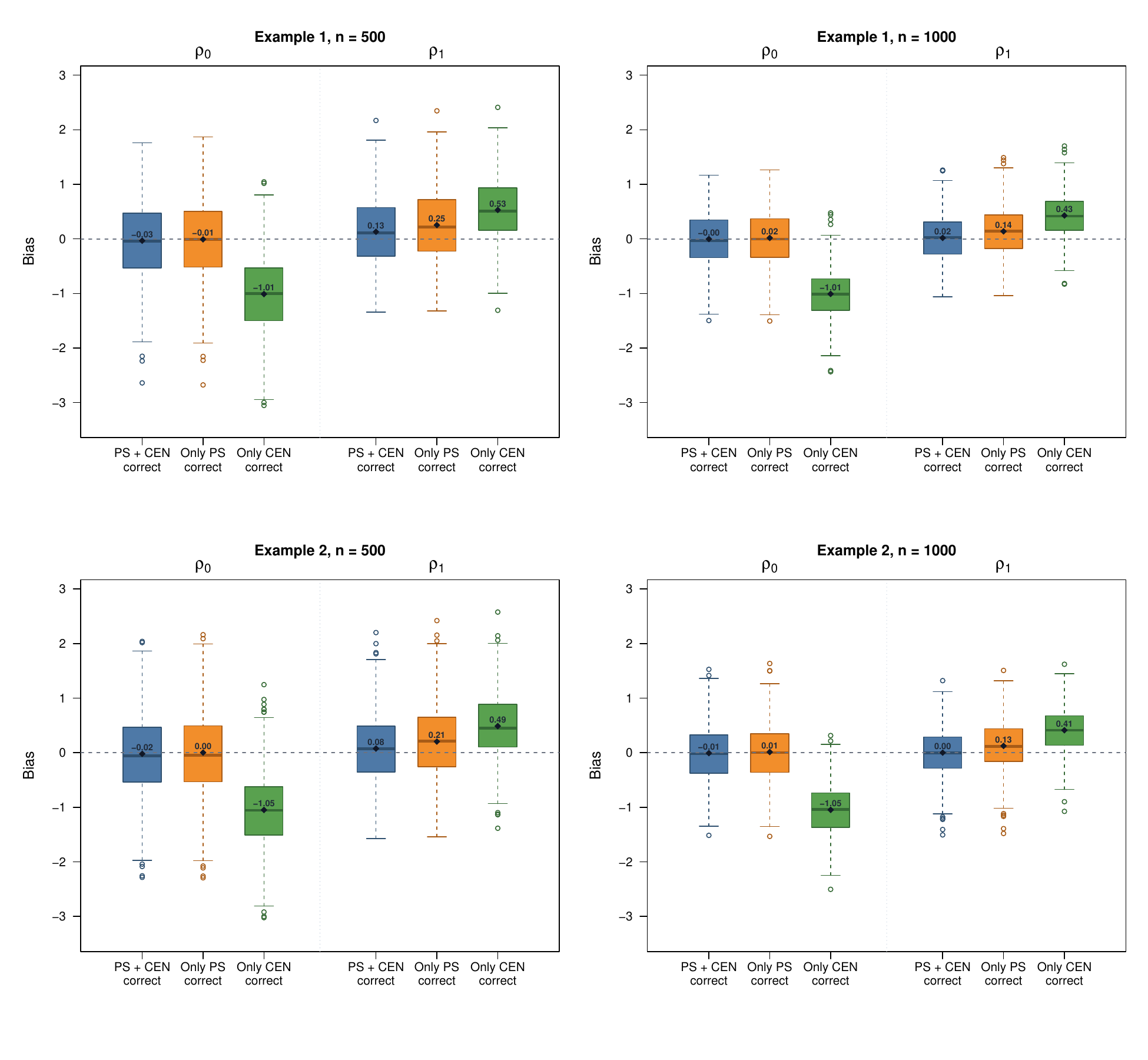}
	\caption{Boxplots of the bias of the doubly robust estimators of \(\rho_0\) and \(\rho_1\) under misspecification of the backward odds-product model $\widetilde{\mathcal M}_{\sop}$. Each boxplot summarizes $500$ Monte Carlo replications. Blue denotes correct specification of both the propensity-score model $\mathcal{M}_{\ps}$ and censoring model $\mathcal{M}_{\cen}$; orange denotes correct specification of only the propensity-score model $\mathcal{M}_{\ps}$; and green denotes correct specification of only the censoring model $\mathcal{M}_{\cen}$. Black diamonds and the accompanying numerical values indicate the mean bias.}
	\label{fig:cir-censoring}
	\par\small\noindent\textit{Alt text: Four panels show boxplots of the biases of the doubly robust estimators of $\rho_0$ and $\rho_1$ for Examples 1 and 2 at sample sizes $500$ and $1000$, with the backward odds-product model misspecified throughout. When both the propensity-score and censoring models are correctly specified, the bias distributions are centered near zero. When only the propensity-score model is correct, the bias remains near zero for $\rho_0$ but is moderately positive for $\rho_1$. When only the censoring model is correct, the bias is consistently about $-1$ for $\rho_0$ and approximately $0.4$--$0.5$ for $\rho_1$. The boxplots generally become narrower as the sample size increases, while the biases under model misspecification persist.}
\end{figure}

To examine more directly the role of the censoring model \(\mathcal M_{\cen}\) in the robustness of the DR estimator, we conduct an additional comparison in which \(\widetilde{\mathcal M}_{\sop}\) is misspecified. Figure~\ref{fig:cir-censoring} displays boxplots of the biases of the DR estimators of \(\rho_0\) and \(\rho_1\) under three configurations: (i) \textsf{PS+CEN correct}: both \(\mathcal{M}_{\ps}\) and \(\mathcal M_{\cen}\) are correctly specified; (ii) \textsf{Only PS correct}: \(\mathcal M_{\ps}\) is correctly specified, but \(\mathcal M_{\cen}\) is misspecified; and (iii) \textsf{Only CEN correct}: \(\mathcal M_{\cen}\) is correctly specified, but \(\mathcal M_{\ps}\) is misspecified. When \(\mathcal M_{\ps}\) is correctly specified, although misspecification of \(\mathcal M_{\cen}\) shows only a comparatively small bias for \(\rho_0\), the bias for \(\rho_1\) is persistently greater than that under correct specification of \(\mathcal M_{\cen}\). This difference for \(\rho_1\) remains apparent as the sample size increases from \(500\) to \(1000\). Conversely, correct specification of \(\mathcal M_{\cen}\) alone does not eliminate bias when \(\mathcal M_{\ps}\) is misspecified. These findings are consistent with Proposition~\ref{prop:sur:DR-CIR}.

\section{Data applications}\label{sec:data}

We illustrate the proposed methods using data from two clinical trials: the prostate cancer trial \citep{byar1980choice} for the survival-ratio framework and the Health Insurance Plan of Greater New York (HIP) Breast Cancer Screening Trial \citep{shapiro1997periodic} for the cumulative-incidence-ratio framework.

\subsection{A randomized trial of prostate cancer}

We analyze a randomized trial among patients with prostate cancer \citep{byar1980choice}. The data were obtained from the Vanderbilt University Department of Biostatistics data repository \citep{hbiostatProstate}. A total of $502$ patients were randomly assigned to placebo or to 0.2, 1.0, or 5.0 mg estrogen, with all-cause mortality as the outcome. One patient with missing age was excluded from all analyses, leaving an analytic sample of $501$ patients. Estrogen therapy may reduce prostate cancer mortality by suppressing testosterone production, but it can also increase cardiovascular and other adverse risks \citep{Turo01022014}. We
then aim to quantify and compare the effects of the three estrogen doses on survival relative to placebo, thereby providing survival-based evidence that may inform dose selection.
\begin{figure}[!t]
	\centering
	\begin{minipage}[t]{0.45\textwidth}
		\centering
		\includegraphics[width=\linewidth]{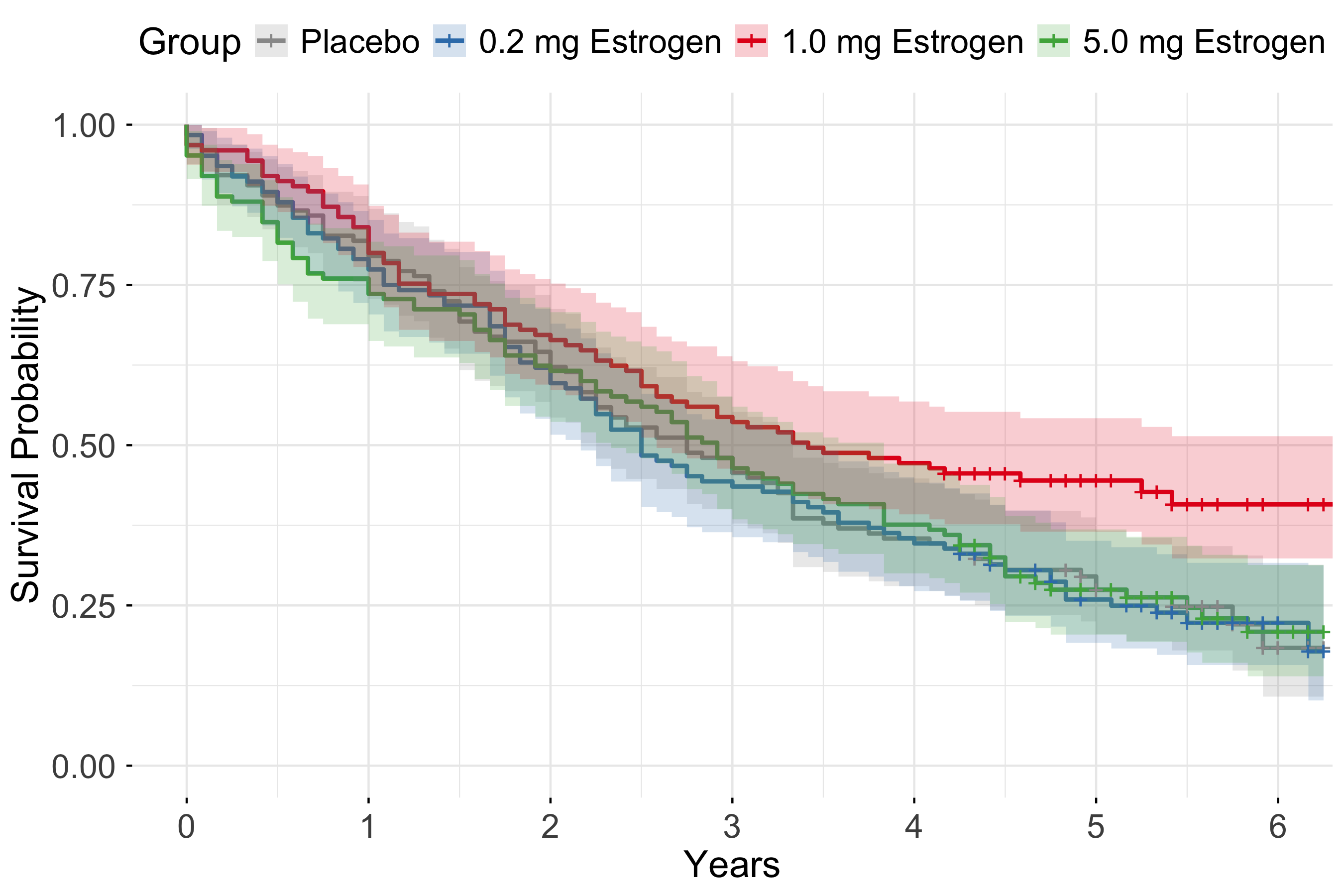}
		\par\smallskip\textbf{(a)}
	\end{minipage}
	\hfill
	\begin{minipage}[t]{0.51\textwidth}
		\centering
		\includegraphics[width=\linewidth]{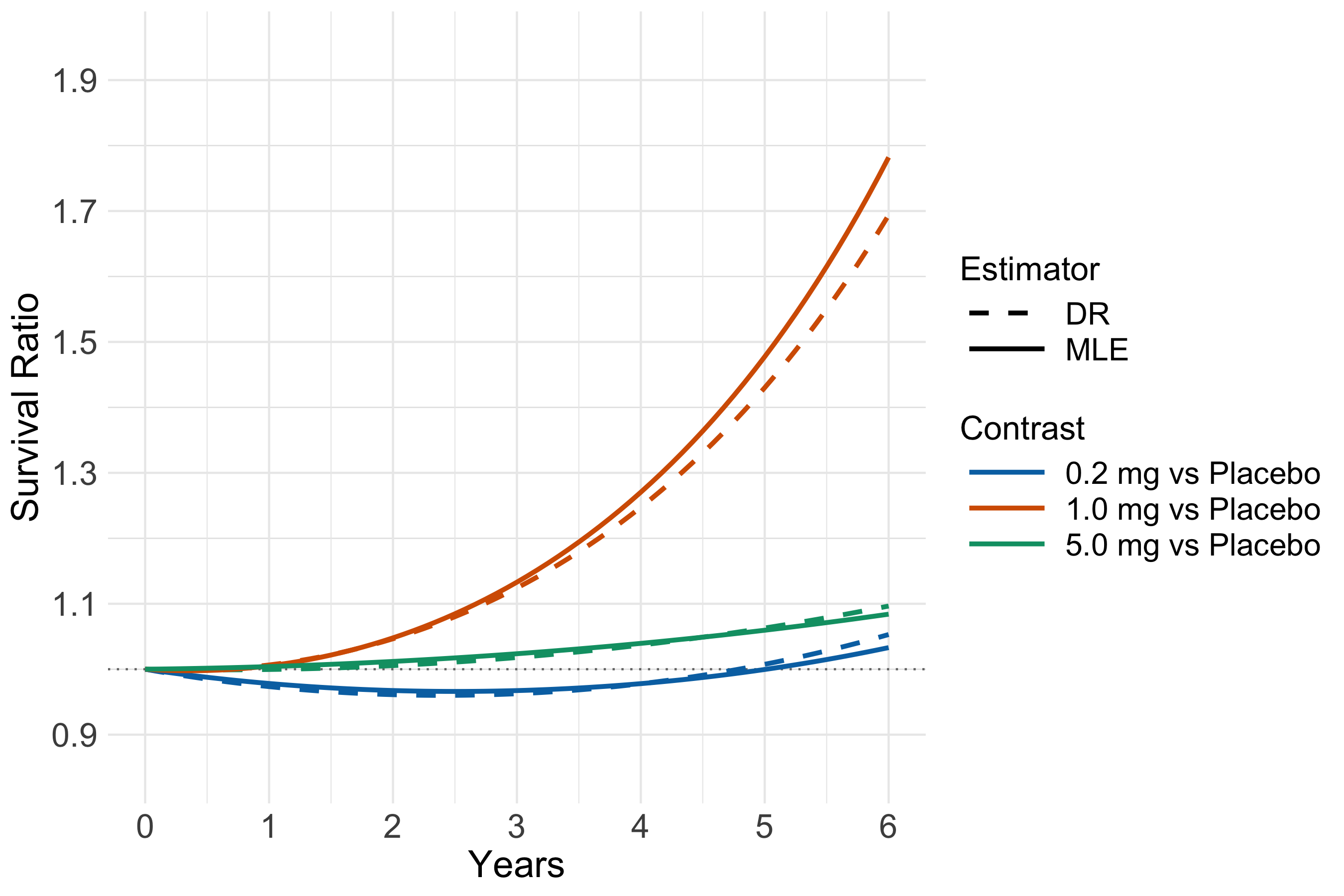}
		\par\smallskip\textbf{(b)}
	\end{minipage}
	\caption{Prostate cancer trial analysis. (a) Kaplan--Meier survival curves for placebo and the three estrogen-dose groups; shaded bands are pointwise 95\% confidence intervals. (b) Estimated survival ratios for each estrogen dose versus placebo. Solid and dashed curves denote the maximum likelihood and doubly robust estimators, respectively; colors identify the estrogen dose.}
	\label{fig:prostate-analysis}
	\par\small\noindent\textit{Alt text: Panel (a) shows Kaplan--Meier survival curves over six years for placebo and estrogen doses of 0.2, 1.0, and 5.0 mg. The curves overlap early; after about three years, estimated survival is highest for the 1.0 mg group, while the other dose groups remain close to placebo. Panel (b) shows solid maximum likelihood and dashed doubly robust fitted survival-ratio curves. The ratios for 0.2 and 5.0 mg remain near one, whereas the ratio for 1.0 mg rises to approximately 1.7--1.8 by six years.}
\end{figure}

Figure \ref{fig:prostate-analysis}(a) displays the estimated marginal survival curves for the four treatment groups using the Kaplan--Meier estimator, and the shaded bands denote the corresponding pointwise 95\% confidence intervals. The estimated survival curves were similar during early follow-up, with substantially overlapping confidence intervals. After approximately 3 years of follow-up, the 1.0 mg estrogen group began to show higher survival probabilities than the other treatment groups, whereas the 0.2 mg and 5.0 mg estrogen groups showed survival probabilities comparable to those of the placebo group.

\begin{table}[!thp]
	\centering
	\caption{Estimates of $\alpha_1$ and $\alpha_2$ using maximum likelihood and doubly robust methods}
	\label{tab:realdata-alpha}
	\small
	\setlength{\tabcolsep}{2.5pt}
	\begin{tabular}{lcrrrr}
		\toprule
		\multirow{2}{*}{Contrast} & \multirow{2}{*}{Parameter}
		& \multicolumn{2}{c}{MLE}
		& \multicolumn{2}{c}{DR} \\
		\cmidrule(lr){3-4}\cmidrule(lr){5-6}
		& & \multicolumn{1}{c}{Estimate (95\% CI)} & \multicolumn{1}{c}{$P$-value}
		& \multicolumn{1}{c}{Estimate (95\% CI)} & \multicolumn{1}{c}{$P$-value} \\
		\midrule
		\multirow{2}{*}{\shortstack[c]{0.2 mg Estrogen\\vs Placebo}}
		& $\alpha_1$ & $-0.028\;(-0.143,\phantom{-}0.088)$ & $0.641$
		& $-0.034\;(-0.150,\phantom{-}0.082)$ & $0.566$ \\
		& $\alpha_2$ & $\phantom{-}0.006\;(-0.016,\phantom{-}0.027)$ & $0.620$
		& $\phantom{-}0.007\;(-0.013,\phantom{-}0.028)$ & $0.496$ \\

		\multirow{2}{*}{\shortstack[c]{1.0 mg Estrogen\\vs Placebo}}
		& $\alpha_1$ & $-0.013\;(-0.115,\phantom{-}0.089)$ & $0.803$
		& $-0.010\;(-0.115,\phantom{-}0.096)$ & $0.855$ \\
		& $\alpha_2$ & $\phantom{-}0.018\;(-0.002,\phantom{-}0.039)$ & $0.078$
		& $\phantom{-}0.016\;(-0.004,\phantom{-}0.037)$ & $0.114$ \\

		\multirow{2}{*}{\shortstack[c]{5.0 mg Estrogen\\vs Placebo}}
		& $\alpha_1$ & $\phantom{-}0.002\;(-0.105,\phantom{-}0.109)$ & $0.969$
		& $-0.004\;(-0.107,\phantom{-}0.100)$ & $0.947$ \\
		& $\alpha_2$ & $\phantom{-}0.002\;(-0.019,\phantom{-}0.023)$ & $0.858$
		& $\phantom{-}0.003\;(-0.016,\phantom{-}0.023)$ & $0.750$ \\
		\bottomrule
	\end{tabular}
\end{table}

We conduct three pairwise analyses, each restricted to the placebo group and one estrogen-dose group, with estrogen coded as \(A=1\) and placebo as \(A=0\). Follow-up is evaluated on the annual grid $(t_1,...,t_6)=(1,...,6)$ years. Let $\alpha=(\alpha_1,\alpha_2)^\top$, and we specify $\Phi(t;X,\alpha)=\alpha_1 t + \alpha_2 t^2$ in $\mathcal{M}_{\sr}$, allowing the log survival ratio to vary nonlinearly over follow-up while satisfying \(\Phi(0;X,\alpha)=0\). $\mathcal{M}_{\sop}$ and $\mathcal{M}_{\ps}$ include age, normal daily activity, cardiovascular disease history, and serum hemoglobin level as covariates $X$. Although $\Phi(t;X,\alpha)$ does not depend on $X$, the conditional survival function $S_a(t;X)$ remains covariate-dependent through $\mathcal{M}_{\sop}$. Table \ref{tab:realdata-alpha} reports maximum likelihood (MLE) and doubly robust (DR) estimates of $\alpha$. Both methods give similar estimates, all 95\% confidence intervals include zero, and the 1.0 mg comparison has the largest positive estimated $\alpha_2$. Consistently, the fitted survival ratios in Figure \ref{fig:prostate-analysis}(b) increase over follow-up for 1.0 mg estrogen but remain near one for 0.2 and 5.0 mg estrogen. The pointwise 95\% confidence intervals for the survival ratio comparing 1.0 mg estrogen with placebo are shown in Supplementary Figure~\ref{supp-fig:prostate-1mg-ci}. At six years, for the 1.0 mg estrogen group, the estimated survival ratio relative to placebo is approximately $1.78$ (95\% CI: $[1.16,\,2.74]$) under maximum likelihood estimation and $1.69$ (95\% CI: $[1.12,\,2.56]$) under doubly robust estimation.

Compared with the marginal Kaplan--Meier curves, the proposed model summarizes the emerging separation of the 1.0 mg group from placebo as a smooth trajectory on the relative-survival scale and provides covariate-adjusted inference for that trajectory. Unlike a hazard-ratio analysis, the resulting contrast directly compares survival probabilities rather than hazards within treatment-specific risk sets. Among the doses examined, the results suggest that 1.0 mg has the strongest estimated improvement in survival over follow-up, although the confidence intervals for the model parameters include zero and the evidence should be interpreted as suggestive rather than conclusive.

Although the estimated survival ratio curve does not vary with covariates \(X\),  substituting the maximum likelihood estimates into \eqref{eq:S0} yields covariate-specific survival functions, complementing the marginal Kaplan--Meier curves in Figure \ref{fig:prostate-analysis}(a) by allowing survival to vary with $X$. Supplementary Figure~\ref{supp-fig:prostate-profile-survival} illustrates this covariate dependence, showing distinct fitted survival curves for two representative baseline health profiles.

\subsection{Health insurance plan study}

We illustrate the proposed cumulative incidence ratio framework using data from the Health Insurance Plan of Greater New York (HIP) Breast Cancer Screening Trial \citep{shapiro1997periodic}. The dataset is available upon approved application through the National Cancer Institute Cancer Data Access System \citep{nciHIPBData}. The original randomized cohort included 60,695 women aged 40--64 years, with 30,130 assigned to the screening-invitation arm and 30,565 to the control arm. Because the proposed framework does not currently accommodate competing risks, we consider all-cause rather than breast-cancer mortality. Following the sample-selection scheme of \citet{yu2015semiparametric}, we restrict the analysis to the 879 women diagnosed with breast cancer within 7 years after randomization and administratively censor follow-up at 10 years. The purpose of this analysis is to illustrate the ability of the proposed models to capture a time-varying cumulative incidence ratio. Because inclusion in this cohort is defined by a post-randomization diagnosis, comparisons between the groups defined by the original random assignment are descriptive and should not be interpreted as intention-to-screen effects in the original randomized population.

\begin{figure}[htbp!]
	\centering
	\begin{minipage}[t]{0.46\textwidth}
		\centering
		\includegraphics[
		width=\linewidth,
		trim=7pt -3pt 3pt 30pt,
		clip
		]{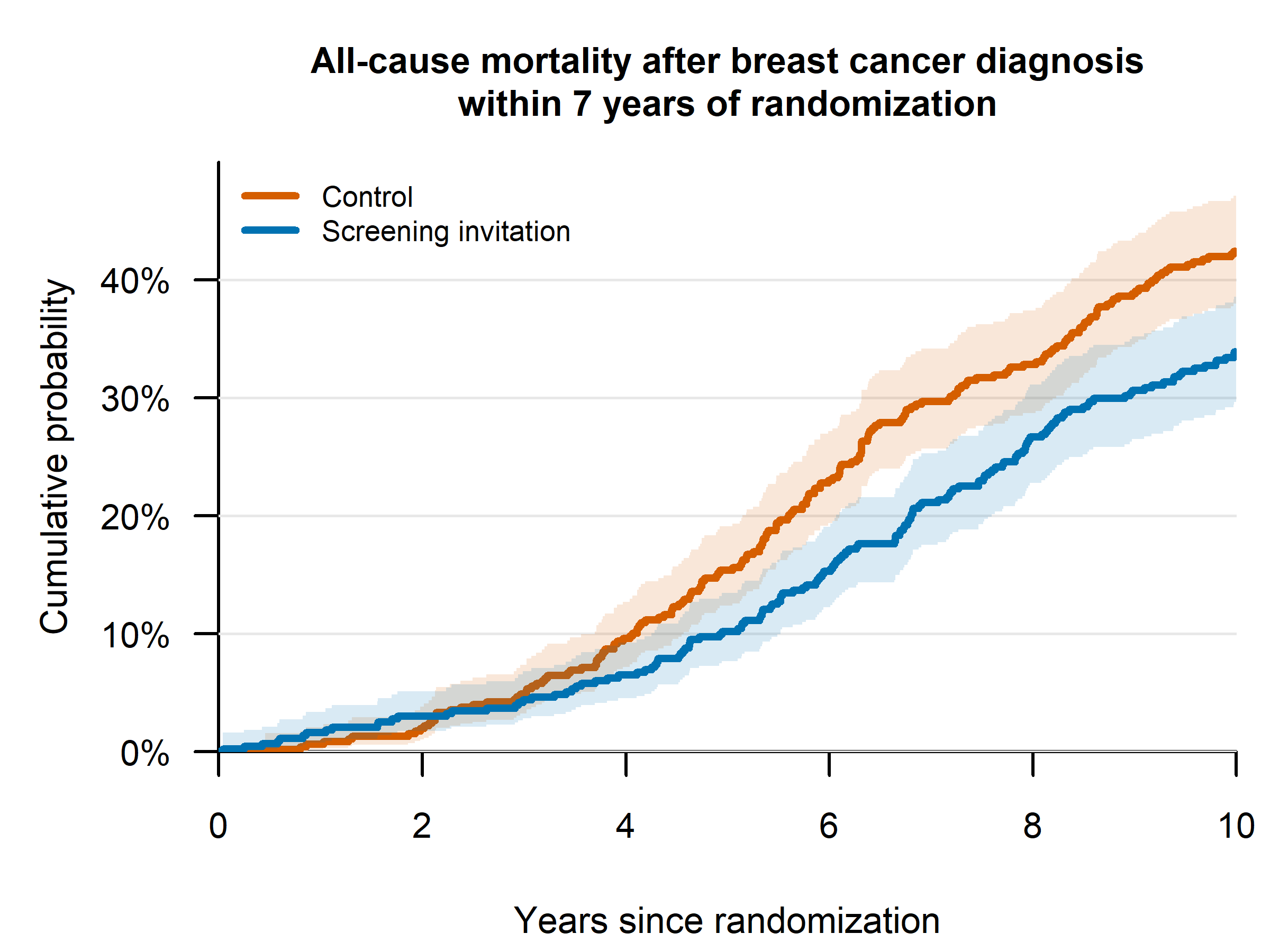}
		\par\smallskip\textbf{(a)}
	\end{minipage}
	\hfill
	\begin{minipage}[t]{0.52\textwidth}
		\centering
		\includegraphics[width=\linewidth]{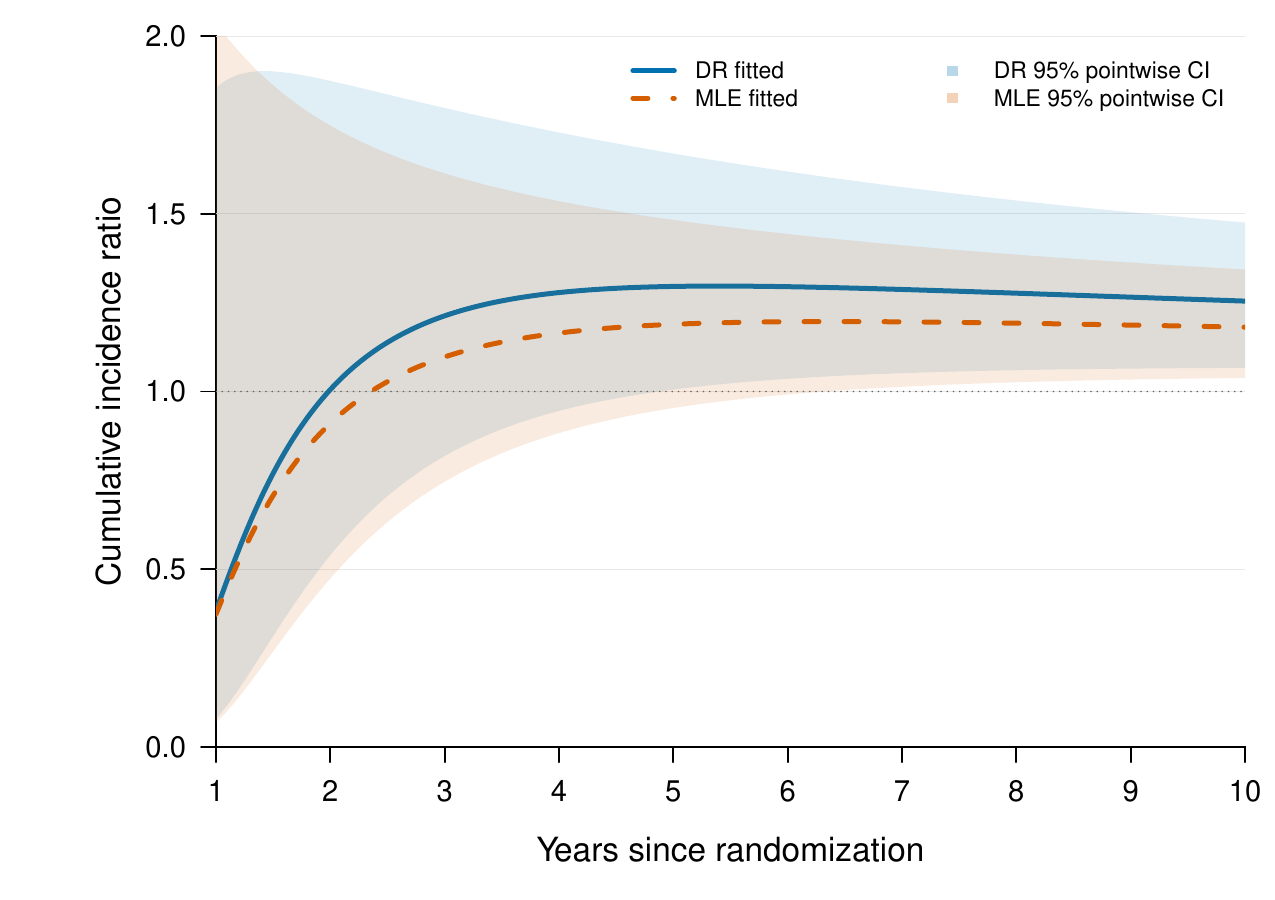}
		\par\smallskip\textbf{(b)}
	\end{minipage}
	\caption{HIP Breast Cancer Screening Trial analysis in the selected cohort. (a) Estimated marginal cumulative incidence curves for the control and screening-invitation groups, calculated as one minus the Kaplan--Meier estimator. (b) Estimated cumulative incidence ratio curves obtained using the maximum likelihood and doubly robust methods, with pointwise 95\% confidence intervals.}
	\label{fig:hip-analysis}
	\par\small\noindent\textit{Alt text: Panel (a) shows marginal cumulative incidence curves for all-cause death over ten years in the selected control and screening-invitation groups. The curves cross early, after which cumulative incidence is lower in the screening-invitation group. Panel (b) shows maximum likelihood and doubly robust fitted cumulative incidence ratio curves with pointwise 95\% confidence bands. The fitted curves cross one after about two years, and the confidence bands are widest early in follow-up.}
\end{figure}

Figure~\ref{fig:hip-analysis}(a) shows the estimated marginal cumulative incidence of all-cause death in the control and screening-invitation groups within this selected cohort over 10 years of follow-up. A notable feature is that the two curves cross during the early follow-up period: the cumulative incidence is initially slightly higher in the screening-invitation group, but the ordering subsequently reverses, with a lower cumulative incidence in the screening-invitation group thereafter. The crossing of the marginal curves motivates a time-varying model that allows the cumulative incidence ratio to cross one. We model the cumulative incidence ratio comparing the control group with the screening-invitation group, with the former coded as \(A=1\). Let $\rho=(\rho_1,\rho_2)^\top$, and we specify
\(\Lambda(t;X,\rho)
=
\frac{\rho_1}{t}
+
\frac{\rho_2\log(t)}{t},
\)
in $\mathcal{M}_{\cir}$. Standardized baseline age is included in the covariate vector $X$ used in $\widetilde{\mathcal M}_{\sop}$ and $\mathcal M_{\ps}$. Thus, although $\Lambda(t;X,\rho)$ does not depend on $X$, the conditional cumulative distribution functions $F_a(t;X)$ remain covariate-dependent through $\widetilde{\mathcal M}_{\sop}$. Because censoring occurs only at the 10-year administrative endpoint, the conditional probability mass function of the censoring time satisfies $g(t_k;X)=0$ for $k=1,\ldots,9$ and $g(t_{10};X)=1$. On the annual time grid $(t_1,\ldots,t_{10})=(1,\ldots,10)$ years, we estimate $\rho$ using the maximum likelihood (MLE) and doubly robust (DR) methods, with the results summarized in Table~\ref{supp-tab:realdata-cir}. The two methods yield similar estimates. The 95\% confidence intervals for $\rho_1$ include zero, whereas those for $\rho_2$ exclude zero under both methods. Figure~\ref{fig:hip-analysis}(b) displays the fitted cumulative incidence ratio curves with their 95\% pointwise confidence intervals.

As shown in Figure~\ref{fig:hip-analysis}(b), the MLE and DR fitted curves exhibit similar temporal patterns, although the fitted DR curve lies consistently above the fitted MLE curve after approximately 2 years. The fitted cumulative incidence ratio crosses one at approximately 2 years for the DR estimator and 2.4 years for the MLE. Among women in this selected cohort, the fitted conditional cumulative incidence of all-cause death was initially lower in the control group than in the screening-invitation group, but the ordering reversed later in follow-up. The pointwise confidence intervals are particularly wide at early time points, reflecting the limited number of early deaths, and become narrower as follow-up increases.

\section{Discussion}\label{sec:discussion}
We developed variation-independent modeling frameworks for time-varying conditional survival and cumulative incidence ratios. By pairing each sequential ratio with an odds-product nuisance model, the proposed parameterizations separate the target parameters from the nuisance parameters while ensuring valid monotone conditional survival and cumulative distribution functions. The survival ratio framework uses a forward construction anchored at the beginning of follow-up, whereas the cumulative incidence ratio framework uses a backward construction. We derived likelihood-based and doubly robust estimators for both frameworks. They can also accommodate the setting \(C\ind T\mid(A,X)\), allowing the censoring distribution to vary across treatment arms.

The choice between the two ratio scales depends on the scientific question and event frequency. The survival ratio is natural when interest centers on the relative probability of remaining event-free and event incidence is sufficiently high that survival probabilities depart meaningfully from one. The cumulative incidence ratio directly compares event probabilities and is therefore more informative for rare events, for which survival ratios can remain close to one even when the relative difference in event risk is substantial. Both frameworks are particularly useful when the ratio of interest is expected to vary over follow-up and a covariate-adjusted model for that trajectory is desired.

Several limitations suggest directions for future work. First, we could extend our framework to accommodate flexible machine-learning methods for nuisance estimation while improving numerical stability. Second, the current method requires a prespecified discrete time grid and a finite-dimensional model for the target ratio. The estimates may be sensitive to the number and locations of the grid points and to the choice of ratio functions. Future work will evaluate this sensitivity, provide practical guidance for selecting the grid and ratio functions, and investigate extensions of the proposed framework to continuous time. Finally, the present development focuses on binary treatments and does not accommodate competing risks. Extensions to continuous treatments and to cause-specific cumulative incidence ratios in the presence of competing events would broaden the range of applications.

\section*{Acknowledgments}

R.Z. was partially supported by a CANSSI Ontario Postdoctoral Fellowship in Statistical Sciences. The authors used ChatGPT to assist with language editing and polishing. The authors reviewed and take full responsibility for the final content. The authors declare no conflicts of interest.

\section*{Data Availability Statement}
The prostate cancer trial data are publicly available from the Vanderbilt University Department of Biostatistics data repository \citep{hbiostatProstate}. The HIP Breast Cancer Screening Trial data are available upon approved application through the National Cancer Institute Cancer Data Access System \citep{nciHIPBData}.

\FloatBarrier
\bibliographystyle{plainnat}
\bibliography{paper-ref}

\clearpage
\appendix
\allowdisplaybreaks[1]
\renewcommand{\thesection}{Appendix \Alph{section}}
\renewcommand{\thesubsection}{\thesection.\arabic{subsection}}
\renewcommand{\theHsection}{supp-appendix-\Alph{section}}
\renewcommand{\theHsubsection}{\theHsection.\arabic{subsection}}
\setcounter{equation}{0}
\setcounter{figure}{0}
\setcounter{table}{0}
\setcounter{theorem}{0}
\setcounter{proposition}{0}
\setcounter{remark}{0}
\renewcommand{\theequation}{S\arabic{equation}}
\renewcommand{\thefigure}{S\arabic{figure}}
\renewcommand{\thetable}{S\arabic{table}}
\renewcommand{\thetheorem}{S\arabic{theorem}}
\renewcommand{\theproposition}{S\arabic{proposition}}
\renewcommand{\theremark}{S\arabic{remark}}
\renewcommand{\theHequation}{supp-equation-\arabic{equation}}
\renewcommand{\theHfigure}{supp-figure-\arabic{figure}}
\renewcommand{\theHtable}{supp-table-\arabic{table}}
\renewcommand{\theHtheorem}{supp-theorem-\arabic{theorem}}
\renewcommand{\theHproposition}{supp-proposition-\arabic{proposition}}
\renewcommand{\theHremark}{supp-remark-\arabic{remark}}

\section*{Supplementary Materials}
Appendices A and B, Tables S1--S3, and Figures S1--S2, which provide additional methodological details, proofs, simulation results, and application figures, are provided below. Code implementing the proposed methods and reproducing the simulation studies and prostate cancer analysis is available in the SR-CIR repository \citep{shi2026srcir}.
\addcontentsline{toc}{section}{Supplementary Materials}

\section{Survival ratio modeling}\label{supp-sec:sr}

\subsection{Maximum likelihood estimation}\label{supp-sec:sr-mle}

The log-likelihood function for an individual is
\begin{align*}
	l_{\sr}(\alpha,\beta)= \sum_{k=1}^{m}  \Big\{ &{\mathbb{I}(Y = t_{k}) \Delta} \log \Big  ( (
	S_{A}(t_{k-1}; X) - S_{A}(t_{k}; X) ) (1 - G(t_{k}; X)) \Big) \\
	&+ {\mathbb{I}(Y = t_{k})(1-\Delta)} \log \Big ( S_{A}(t_{k}; X) g(t_{k}; X) \Big )  \Big\},
\end{align*}
where $G(t_k;X)= \sum_{i=1}^{k-1}g(t_i;X)$. Let the maximum likelihood estimators $\widehat{\alpha}$ and $\widehat{\beta}$ solve
\begin{align*}
	0=\mathbb{P}_n  [ S_{\sr}(\alpha,\beta)],
\end{align*}
where $S_{\sr}(\alpha,\beta)=\frac{\partial}{\partial (\alpha^\top,\beta^\top)^\top} l_{\sr}(\alpha,\beta)$ is the score function for $\alpha$ and $\beta$, and $\mathbb{P}_n$ denotes the empirical average. Inference for \(\alpha\) and \(\beta\) is based on the sandwich variance estimator \citep{white1982maximum}, and the corresponding Wald confidence intervals are constructed.

\subsection{Doubly robust estimation}\label{supp-sec:sr-dr}

\begin{proof}[Proof of Theorem \ref{thm:sur:effscore}]
For convenience, we omit $X$ in the following notation, and in particular let $S_A(t)=S_A(t;X)$, $f_{\sr}(t)=f_{\sr}(t;X)$, $G(t;X)=G(t)$ and $g(t;X)=g(t)$.

The survival ratio model $\mathcal{M}_{\sr}$ can be given by
\begin{align*}
\log S_A(t) = \log S_0(t) + \alpha^\top f_{\sr}(t)A,
\end{align*}
and we use the following six steps to derive the efficient score $S^*(\alpha)$ for $\alpha$ under model $\mathcal{M}_{\sr}$.

\noindent\textbf{Step 1: (compute the likelihood and score function)}
By Assumption of $C \ind (A,T)|X$, the likelihood is
\begin{align*}
&P(Y,\Delta,A,X) = P(Y,\Delta|A,X) P(A,X)\\
=& \prod_{k=1}^{m} \Big \{ [(S_A(t_{k-1})-S_A(t_{k}))(1-G(t_{k}))]^{\mathbb{I}(Y=t_{k})\Delta}  [S_A(t_{k})g(t_{k})]^{\mathbb{I}(Y=t_{k})(1-\Delta)}\Big \} \times P(A,X)\\
=& \prod_{k=1}^m \Big \{ \Big ( 1- \frac{S_A(t_k)}{S_A(t_{k-1})} \Big )^{\mathbb{I}(Y=t_k)\prod_{j=0}^{k-1}(1- \mathbb{I}(Y=t_j))\Delta} \Big ( \frac{S_A(t_k)}{S_A(t_{k-1})} \Big )^{(1-\mathbb{I}(Y=t_k)\Delta)\prod_{j=0}^{k-1}(1- \mathbb{I}(Y=t_j))}  \Big \}\\*
& \times \prod_{k=1}^{m} \Big \{ \Big(1-G(t_{k})\Big)^{\mathbb{I}(Y=t_{k})\Delta} g(t_{k})^{\mathbb{I}(Y=t_{k})(1-\Delta)}\Big \}\times P(A,X),
\end{align*}
and we have
\begin{align*}
&\log P(Y,\Delta,A,X) \\
=&  \sum_{k=1}^m \prod_{j=0}^{k-1}(1- \mathbb{I}(Y=t_j))\Big [ \mathbb{I}(Y=t_k)\Delta \log \Big( 1- \frac{S_A(t_k)}{S_A(t_{k-1})} \Big)\\*
&\qquad {}+  (1-\mathbb{I}(Y=t_k)\Delta)\log \Big( \frac{S_A(t_k)}{S_A(t_{k-1})} \Big)\Big]\\
&+ \sum_{k=1}^m \Big \{ \mathbb{I}(Y=t_k) \Delta \log \Big ( 1- G(t_k) \Big ) + \mathbb{I}(Y=t_k) (1-\Delta) \log(  g(t_k)  )\Big \}\\*
&\qquad {}+ \log P(A,X).
\end{align*}
Under model $\mathcal{M}_{\sr}$, we compute the score function
\begin{align*}
S_\alpha &= \frac{\partial \log P(Y,\Delta,A,X)}{\partial \alpha}\\
=& A \sum_{k=1}^m \prod_{j=0}^{k-1}(1- \mathbb{I}(Y=t_j))\Big [ \mathbb{I}(Y=t_k)\Delta \frac{ - \frac{S_A(t_k)}{S_A(t_{k-1})} (f_{\sr}(t_k)-f_{\sr}(t_{k-1}))}{ 1- \frac{S_A(t_k)}{S_A(t_{k-1})}}\\*
&\qquad {}+  (1-\mathbb{I}(Y=t_k)\Delta) (f_{\sr}(t_k)-f_{\sr}(t_{k-1}))\Big]\\
=& A \sum_{k=1}^m  (f_{\sr}(t_k)-f_{\sr}(t_{k-1})) \prod_{j=0}^{k-1}(1- \mathbb{I}(Y=t_j)) \Big [ \frac{1- \mathbb{I}(Y=t_k)\Delta - \frac{S_A(t_k)}{S_A(t_{k-1})}}{1- \frac{S_A(t_k)}{S_A(t_{k-1})}}\Big]\\
=& A \sum_{k=1}^m \Big \{ (f_{\sr}(t_k)-f_{\sr}(t_{k-1})) \frac{S_A(t_k)}{1 - \frac{S_A(t_k)}{S_A(t_{k-1})}} \Big [ \frac{1}{S_0(t_k)} H_k(\alpha)\\*
&\qquad {}+ \frac{1}{S_0(t_k)} C_k(\alpha) - \frac{1}{S_0(t_{k-1})} H_{k-1}(\alpha)\Big] \Big \}\\
=& A \mathcal{F}_{\sr,X}^\top \Gamma Q_{A,X} R(\alpha),
\end{align*}
where $p_\alpha=\dim(\alpha)$, $\mathcal{F}_{\sr,X} = (f_{\sr}(t_1),...,f_{\sr}(t_m))^\top \in \mathbb{R}^{m\times p_\alpha}$, and $\Gamma =(\gamma_{i,j})_{m\times m}$ with
\begin{align*}
\gamma_{ij}=
\begin{cases}
1 & \text{ if }  j=i\\
-1 & \text{ if }  j=i+1\\
0 & \text{ otherwise},
\end{cases}
\end{align*}
$Q_{A,X}= \text{diag}(q_{A,X}^1,...,q_{A,X}^m)$ with $q_{A,X}^k=\frac{S_A(t_k)}{1 - \frac{S_A(t_k)}{S_A(t_{k-1})}}$, and $R(\alpha) = (R_1(\alpha),...,R_m(\alpha))^\top$ with
\begin{align*}
&R_k(\alpha)=\frac{1}{S_0(t_k)} H_k(\alpha) + \frac{1}{S_0(t_k)} C_k(\alpha) - \frac{1}{S_0(t_{k-1})} H_{k-1}(\alpha),\\
&H_k(\alpha) = \frac{S_0(t_k)}{S_A(t_k)}\prod_{j=0}^{k} (1-\mathbb{I}(Y=t_j)) = \exp(-\alpha^\top f_{\sr}(t_k)A) \prod_{j=0}^{k} (1-\mathbb{I}(Y=t_j)),\\
&C_k(\alpha) =\frac{S_0(t_k)}{S_A(t_k)}(1-\Delta)\mathbb{I}(Y=t_k) = \exp(-\alpha^\top f_{\sr}(t_k)A)(1-\Delta)\mathbb{I}(Y=t_k).
\end{align*}

\noindent\textbf{Step 2: (Nuisance tangent space)}
To construct the nuisance tangent space, we define an arbitrary parametric submodel by substituting $S_0(t;\zeta_1)$, $g(t;\zeta_2)$ and $P(A,X;\zeta_3)$ into $P(Y,\Delta,A,X)$, where $\zeta_1$, $\zeta_2$ and $\zeta_3$ are finite-dimensional nuisance parameters of dimensions $r_1$, $r_2$ and $r_3$, respectively. Since $\zeta_1$, $\zeta_2$, and $\zeta_3$ are variation independent, they can be separated in $\log P(Y,\Delta,A,X)$. Then, the nuisance tangent space $\Lambda$ can be written as a direct sum of three orthogonal spaces
\begin{align*}
\Lambda = \Lambda_{1s} \oplus \Lambda_{2s} \oplus \Lambda_{3s},
\end{align*}
where
\[
\Lambda_{js} = \text{mean-square closure of }\{B S_{\zeta_j}(Y,\Delta,A,X) \text{ for all }B\},
\]
and $S_{\zeta_j}(Y,\Delta,A,X)=\partial \log P(Y,\Delta,A,X)/ \partial \zeta_j$ for $j=1,2,3$. Let $\mathcal{H}$ denote the tangent space of the enlarged model in which $S_0(t)$ and $S_1(t)$ are unrestricted. Then, this tangent space admits the following orthogonal decomposition:
\begin{align*}
\mathcal{H} = \Lambda_{1s}^* \oplus \Lambda_{2s} \oplus \Lambda_{3s},
\end{align*}
where $\Lambda_{1s}^*$ is the space associated with $S_A(t)$ and $S_A(t)$ is now left arbitrary. Throughout this proof, $\Lambda^{\perp}$ denotes the orthogonal complement of $\Lambda$ within $\mathcal{H}$.

\noindent\textbf{Step 3: (Compute $\Lambda_{1s}$)}
In order to derive the space $\Lambda_{1s}$, we consider the parametric submodel
\begin{align*}
\log S_{ 0}(t;\zeta_1) = \log S_0(t) + \zeta_1^\top h(t)
\end{align*}
for an arbitrary function $h(t)$, which implies
\begin{align*}
\log S_{A}(t;\zeta_1) = \log S_0(t) + \zeta_1^\top h(t) + \alpha^\top f_{\sr}(t)A.
\end{align*}
Then, we have the score function with respect to $\zeta_1$
\begin{align*}
S_{\zeta_1}=& \sum_{k=1}^m \Big \{ (h(t_k)-h(t_{k-1})) \frac{S_A(t_k)}{1 - \frac{S_A(t_k)}{S_A(t_{k-1})}} \Big [ \frac{1}{S_0(t_k)} H_k(\alpha) + \frac{1}{S_0(t_k)} C_k(\alpha) - \frac{1}{S_0(t_{k-1})} H_{k-1}(\alpha)\Big] \Big \}\\
=& \mathcal{H}_X^\top \Gamma Q_{A,X} R(\alpha),
\end{align*}
where $\mathcal{H}_X = (h(t_1)^\top,..., h(t_m)^\top)^\top\in \mathbb{R}^{m\times r_1}$. Thus,
\begin{align*}
\Lambda_{1s} = \Big \{ \mathcal{H}_X^\top \Gamma Q_{A,X} R(\alpha) : \mathcal{H}_X \Big \}.
\end{align*}

\noindent\textbf{Step 4: (Compute $\Lambda_{1s}^*$)}
In order to derive the space $\Lambda_{1s}^*$, we consider the parametric submodel
\begin{align*}
\log S_{ A}(t;\zeta_1^*) = \log S_A(t) + (\zeta_1^*)^\top b_A(t)
\end{align*}
for an arbitrary function $b_A(t)$.
Then, we have
\begin{align*}
S_{\zeta_1^*} =& \sum_{k=1}^m \Big \{ (b_A(t_k)-b_A(t_{k-1})) \frac{S_A(t_k)}{1 - \frac{S_A(t_k)}{S_A(t_{k-1})}} \Big [ \frac{1}{S_0(t_k)} H_k(\alpha)\\*
&\qquad {}+ \frac{1}{S_0(t_k)} C_k(\alpha) - \frac{1}{S_0(t_{k-1})} H_{k-1}(\alpha)\Big] \Big \}\\
=& \mathcal{B}_{A,X}^\top \Gamma Q_{A,X} R(\alpha),
\end{align*}
where $\mathcal{B}_{A,X} = (b_A(t_1)^\top,..., b_A(t_m)^\top)^\top \in \mathbb{R}^{m\times r_1}$. Thus,
\begin{align*}
\Lambda_{1s}^* = \Big \{ \mathcal{B}_{A,X}^\top \Gamma Q_{A,X} R(\alpha) : \mathcal{B}_{A,X} \Big \}.
\end{align*}

\noindent\textbf{Step 5: (Compute $\Lambda^{\perp}$)}
We find that $\Lambda^{\perp}\subset \Lambda_{1s}^*$ and that every element of $\Lambda^{\perp}$ must be orthogonal to $\Lambda_{1s}$. To compute $\Lambda^{\perp}$, it suffices to take an arbitrary element in $\Lambda_{1s}^*$ and find its residual after projecting it onto every element in $\Lambda_{1s}$. In particular, we seek $\mathcal{H}_X^*$ such that
\begin{align}\label{supp-eq:proj}
\mathbb{E}\Big[(\mathcal{B}_{A,X}^\top J_{A,X} - (\mathcal{H}^*_X)^\top J_{A,X})^\top \mathcal{H}_X^\top J_{A,X}\Big]=0
\end{align}
for all $\mathcal{H}_X$, where $J_{A,X}=\Gamma Q_{A,X} R(\alpha)$. Following the argument on page 17 of \citet{tsiatis2006semiparametric}, the condition in \eqref{supp-eq:proj} for all $\mathcal{H}_X$ is equivalent to
\begin{align}\label{supp-eq:proj2}
\mathbb{E}\Big ( \mathcal{V}_{A,X} (J_{A,X})^\top \mid X  \Big) = \mathbf{0}_{r_1 \times m},
\end{align}
where $\mathcal{V}_{A,X}= \mathcal{B}_{A,X}^\top J_{A,X} - (\mathcal{H}^*_X)^\top J_{A,X}$.

It follows from \eqref{supp-eq:proj} that we have
\begin{align*}
0&=\mathbb{E}[\mathcal{V}_{A,X}^\top \mathcal{H}_X^\top J_{A,X}]= \mathbb{E}\Big[\sum_{i,j}(\mathcal{V}_{A,X})_i (J_{A,X})_j (\mathcal{H}_X)_{ji}\Big]\\
&=\mathbb{E}\Big[\sum_{i,j} \mathbb{E}\Big ((\mathcal{V}_{A,X})_i (J_{A,X})_j \mid X \Big ) (\mathcal{H}_X)_{ji}\Big]
\end{align*}
for any $(\mathcal{H}_X)_{ji}$, which means that
\begin{align*}
\mathbb{E}\Big ( (\mathcal{V}_{A,X})_i (J_{A,X})_j \mid X  \Big) = 0
\end{align*}
for any $i,j$.
Thus, $\mathbb{E}\Big ( \mathcal{V}_{A,X} (J_{A,X})^\top \mid X  \Big) = \mathbf{0}$. Conversely, if $\mathbb{E}\Big ( \mathcal{V}_{A,X} (J_{A,X})^\top \mid X  \Big) = \mathbf{0}$ holds, for any $\mathcal{H}_X$, we have \eqref{supp-eq:proj} being true.

Then, it follows from \eqref{supp-eq:proj2} that
\begin{align*}
\mathbf{0} &= \mathbb{E}\Big ( (\mathcal{B}_{A,X}^\top J_{A,X} - (\mathcal{H}^*_X)^\top J_{A,X})(J_{A,X})^\top \mid X  \Big)\\
&=\mathbb{E}\Big ( (\mathcal{B}_{A,X}  - \mathcal{H}^*_X )^\top J_{A,X} (J_{A,X})^\top \mid X  \Big)\\
&=\mathbb{E}\Big ( \mathcal{B}_{A,X}^\top \Sigma_{A,X}   - (\mathcal{H}^*_X)^\top \Sigma_{A,X} \mid X  \Big)\\
&= \mathbb{E}(\mathcal{B}_{A,X}^\top \Sigma_{A,X} \mid X) - (\mathcal{H}^*_X)^\top \mathbb{E}(\Sigma_{A,X} \mid X) ,
\end{align*}
where $\Sigma_{A,X}=\mathbb{E}(J_{A,X} (J_{A,X})^\top \mid A,X)$.
Hence, we obtain
\begin{align*}
\mathcal{H}^*_X =  \Big [\mathbb{E}(\Sigma_{A,X} \mid X) \Big ]^{-1}  \mathbb{E}(\Sigma_{A,X} \mathcal{B}_{A,X} \mid X),
\end{align*}
and finally
\begin{align*}
\Lambda^\perp = \Big \{ (\mathcal{B}_{A,X} - \mathcal{H}_X^* )^\top J_{A,X} : \mathcal{B}_{A,X} \Big \}.
\end{align*}

\noindent\textbf{Step 6: (Compute the efficient score)}
It follows from $J_{A,X}=\Gamma Q_{A,X} R(\alpha)$ that we have
\begin{align*}
\Sigma_{A,X} =& \mathbb{E}(J_{A,X} J_{A,X}^\top \mid A,X)\\
=& \mathbb{E}(\Gamma Q_{A,X}R(\alpha) R(\alpha)^\top  Q_{A,X} \Gamma^\top\mid A,X)\\
=& \Gamma Q_{A,X} \mathbb{E}(R(\alpha) R(\alpha)^\top \mid A,X)   Q_{A,X} \Gamma^\top\\
=& \Gamma \widetilde{\Sigma}_{A,X} \Gamma^\top,
\end{align*}
where $\widetilde{\Sigma}_{A,X}$ is a diagonal matrix with diagonal element $\widetilde{\sigma}_{A,X}^k=q_{A,X}^k (1-G(t_k))$.

Hence, the efficient score for $\alpha$ is
\begin{align*}
S^*({\alpha})
&=\mathcal{F}_{\sr,X}^\top \Big (A \mathbf{I}_m   - \mathbb{E}(A\Sigma_{A,X} \mid X) \Big [\mathbb{E}(\Sigma_{A,X} \mid X) \Big ]^{-1}  \Big ) J_{A,X}\\
&=\mathcal{F}_{\sr,X}^\top \Gamma \Big (A \mathbf{I}_m   - \mathbb{E}(A  \widetilde{\Sigma}_{A,X} \mid X) \Big [ \mathbb{E}(\widetilde{\Sigma}_{A,X}  \mid X) \Big ]^{-1} \Big )Q_{A,X} R(\alpha)\\
&=\mathcal{F}_{\sr,X}^\top \Gamma \Big (A \mathbf{I}_m   - \mathbb{E}(A  Q_{A,X} \mid X) \Big [ \mathbb{E}(Q_{A,X}  \mid X) \Big ]^{-1} \Big )Q_{A,X} R(\alpha).
\end{align*}
Since $\mathbb{E}(A  q^k_{A,X} \mid X) = q^k_{1,X} e(X)$
and $\mathbb{E}( q^k_{A,X} \mid X) = q^k_{1,X} e(X) + q^k_{0,X}(1-e(X))$,
we have
\begin{align*}
&q_{A,X}^k(A-\frac{\mathbb{E}(A  q^k_{A,X} \mid X)}{\mathbb{E}(q^k_{A,X} \mid X)})\\
=&(Aq_{1,X}^k + (1-A)q_{0,X}^k)(A-\frac{\mathbb{E}(A  q^k_{A,X} \mid X)}{\mathbb{E}(q^k_{A,X} \mid X)})\\
=&Aq_{1,X}^k(A-\frac{\mathbb{E}(A  q^k_{A,X} \mid X)}{\mathbb{E}(q^k_{A,X} \mid X)}) + (1-A)q_{0,X}^k (A-\frac{\mathbb{E}(A  q^k_{A,X} \mid X)}{\mathbb{E}(q^k_{A,X} \mid X)})\\
=&Aq_{1,X}^k(1-\frac{\mathbb{E}(A  q^k_{A,X} \mid X)}{\mathbb{E}(q^k_{A,X} \mid X)}) - (1-A)q_{0,X}^k \frac{\mathbb{E}(A  q^k_{A,X} \mid X)}{\mathbb{E}(q^k_{A,X} \mid X)}\\
=& \frac{Aq_{1,X}^k \mathbb{E}(q^k_{A,X} \mid X)-Aq_{1,X}^k \mathbb{E}(Aq^k_{A,X} \mid X) - (1-A)q_{0,X}^k \mathbb{E}(Aq^k_{A,X} \mid X) }{\mathbb{E}(q^k_{A,X} \mid X)}\\
=& \frac{Aq_{1,X}^k (q^k_{1,X} e(X) + q^k_{0,X}(1-e(X)))-A(q_{1,X}^k)^2 e(X) - (1-A)q_{0,X}^k q_{1,X}^k e(X)}{\mathbb{E}(q^k_{A,X} \mid X)}\\
=&\frac{(A-e(X))q_{0,X}^k q_{1,X}^k}{\mathbb{E}(q^k_{A,X} \mid X)}\\
=&\frac{q_{0,X}^k}{e(X)}\frac{\mathbb{E}(A q^k_{A,X} \mid X)}{\mathbb{E}(q^k_{A,X} \mid X)}(A-e(X)).
\end{align*}
Thus, it holds true that
\begin{align*}
&\Big (A \mathbf{I}_m   -  \mathbb{E}(A  Q_{A,X} \mid X) \Big [\mathbb{E}(  Q_{A,X} \mid X) \Big ]^{-1}  \Big ) Q_{A,X}\\*
={}& (A-e(X)) D_X \mathbb{E}(A  Q_{A,X} \mid X) \Big [\mathbb{E}(  Q_{A,X} \mid X) \Big ]^{-1},
\end{align*}
where $D_X=\text{diag}(d_X^1,...,d_X^m)$ with $d_X^k=\frac{q_{0,X}^k}{e(X)} = \frac{S_0(t_k)}{e(X)(1-\frac{S_0(t_k)}{S_0(t_{k-1})})}$.

Hence, the efficient score is
\begin{align*}
S^*(\alpha) &= \mathcal{F}_{\sr,X}^\top \Gamma \Big (A \mathbf{I}_m   - \mathbb{E}(A  Q_{A,X} \mid X) \Big [ \mathbb{E}(Q_{A,X}  \mid X) \Big ]^{-1} \Big )Q_{A,X} R(\alpha)\\
&=\mathcal{W}_{\text{eff}}(X)~R(\alpha)(A-e(X)),
\end{align*}
where $\mathcal{W}_{\text{eff}}(X)=\mathcal{F}_{\sr,X}^\top \Gamma  D_X \mathbb{E}(A  Q_{A,X} \mid X) \Big [\mathbb{E}(  Q_{A,X} \mid X) \Big ]^{-1}$.
\end{proof}

Under this logistic working model, let $\widehat{\gamma}$ solve the following equation
\begin{align*}
	0=\mathbb{P}_n [X\{A-e(X;\gamma)\}],
\end{align*}
and the doubly robust estimator $\widehat{\alpha}_{\DR}(\mathcal{W})$ is obtained by solving equation \eqref{eq:effscore}.

\begin{remark}(Efficient weighting matrix $\mathcal{W}_{\eff}(X; \widehat{\alpha}, \widehat{\beta},\widehat{\gamma})$) Let $\mathcal{W}_{\eff}(X; \widehat{\alpha}, \widehat{\beta},\widehat{\gamma})$ denote the plug-in version of $\mathcal{W}_{\eff}(X)$, obtained by replacing $S_a(t_k;X)$ with $S_a(t_k;X,\widehat\alpha,
	\widehat\beta)$ and replacing $e(X)$ with $e(X;\widehat\gamma)$, for $k=1,\ldots,m$ and $a=0,1$. Then, by setting $\mathcal{W}(X)=\mathcal{W}_{\eff}(X; \widehat{\alpha}, \widehat{\beta},\widehat{\gamma})$, we obtain the estimator $\widehat{\alpha}_{\DR}(\mathcal{W}_{\eff})$.
\end{remark}

\begin{proof}[Proof of Proposition \ref{prop:sur:dr}]
For convenience, we omit $X$ in the following notation, and in particular let $S_A(t)=S_A(t;X)$, $f_{\sr}(t)=f_{\sr}(t;X)$, $G(t;X)=G(t)$ and $g(t;X)=g(t)$. We also define \(G(t_{m+1})=\sum_{i=1}^{m}g(t_i)=1\), since \(C\) takes values in \(\{t_1,\ldots,t_m\}\).

It follows from Slutsky's theorem that the probability limit of $U_{\sr}(\alpha;\widehat{\alpha},\widehat{\beta},\widehat{\gamma};\mathcal{W})$ is
\begin{align*}
U_{\sr}(\alpha;\alpha^*,\beta^*,\gamma^*;\mathcal{W}):= \mathcal{W}(X)R(\alpha;{\alpha}^*,{\beta}^*)(A-e(X;{\gamma}^*)),
\end{align*}
where ${\alpha}^*,{\beta}^*$ and ${\gamma}^*$ are the probability limits of $\widehat{\alpha},\widehat{\beta}$ and $\widehat{\gamma}$, respectively.
Let $\dot{\alpha}$, $\dot{\beta}$ and $\dot{\gamma}$ be the unknown true values of $\alpha$, $\beta$ and $\gamma$, respectively. When the model $\mathcal{M}_{\sop}$ is correctly specified, $(\alpha^*,\beta^*)=(\dot{\alpha},\dot{\beta})$; when the model $\mathcal{M}_{\ps}$ is correctly specified, $\gamma^*=\dot{\gamma}$.

Firstly, to show that \eqref{eq:effscore} is an unbiased estimating equation for $\alpha$ under $\mathcal{M}_{\union}$, it suffices to verify that $\mathbb{E}[U_{\sr}(\dot{\alpha};\alpha^*,\beta^*,\gamma^*;\mathcal{W})]=0$ under $\mathcal{M}_{\union}$. Recall that
\begin{align*}
&\mathbb{E}[(A-e(X))R_k(\alpha)|X]\\*
={}& \mathbb{E}\Big\{(A-e(X))\Big[ \frac{1}{S_0(t_k)} H_k(\alpha) + \frac{1}{S_0(t_k)} C_k(\alpha)\\*
&\qquad {}- \frac{1}{S_0(t_{k-1})} H_{k-1}(\alpha)\Big]\Big|X \Big \},
\end{align*}
and we have
\begin{align*}
&\mathbb{E}(AH_k(\dot{\alpha}) | X)= \mathbb{E}(A \exp(-\dot{\alpha}^\top f_{\sr}(t_k)A) \prod_{j=0}^k (1-\mathbb{I}(Y=t_j)) | X) = S_0(t_k)(1-G(t_{k+1})) e(X),\\
& \mathbb{E}(H_k(\dot{\alpha}) | X) = \mathbb{E}( \exp(-\dot{\alpha}^\top f_{\sr}(t_k)A) \prod_{j=0}^k (1-\mathbb{I}(Y=t_j)) | X) = S_0(t_k)(1-G(t_{k+1})),\\
&\mathbb{E}(AC_k(\dot{\alpha}) | X)= \mathbb{E}(A \exp(-\dot{\alpha}^\top f_{\sr}(t_k)A) (1-\Delta)\mathbb{I}(Y=t_k) | X) = S_0(t_k)g(t_k) e(X),\\
&\mathbb{E}(C_k(\dot{\alpha}) | X)= \mathbb{E}(\exp(-\dot{\alpha}^\top f_{\sr}(t_k)A) (1-\Delta)\mathbb{I}(Y=t_k) | X) = S_0(t_k)g(t_k).
\end{align*}

If $\mathcal{M}_{\sop}$ is correctly specified, then $(\alpha^*,\beta^*)=(\dot{\alpha},\dot{\beta})$ and $S_0(t)=S_0(t;\dot{\alpha},\dot{\beta})$. Hence,
\begin{align*}
&\mathbb{E}[(A-e(X))R_k(\alpha)|X] \\
=& \mathbb{E}\Big\{(A-e(X;\gamma^*))\Big[ \frac{1}{S_0(t_k;\dot{\alpha},\dot{\beta})} H_k(\dot{\alpha}) + \frac{1}{S_0(t_k;\dot{\alpha},\dot{\beta})} C_k(\dot{\alpha}) - \frac{1}{S_0(t_{k-1};\dot{\alpha},\dot{\beta})} H_{k-1}(\dot{\alpha})\Big]\Big|X \Big \}\\
=&\frac{1}{S_0(t_k;\dot{\alpha},\dot{\beta})} \Big[ \mathbb{E}(AH_k(\dot{\alpha})|X) + \mathbb{E}(AC_k(\dot{\alpha})|X) \Big] - \frac{1}{S_0(t_{k-1};\dot{\alpha},\dot{\beta})} \mathbb{E}(AH_{k-1}(\dot{\alpha})|X)\\
&-e(X;\gamma^*) \mathbb{E}\Big[ \frac{1}{S_0(t_k;\dot{\alpha},\dot{\beta})} H_k(\dot{\alpha}) + \frac{1}{S_0(t_k;\dot{\alpha},\dot{\beta})} C_k(\dot{\alpha}) - \frac{1}{S_0(t_{k-1};\dot{\alpha},\dot{\beta})} H_{k-1}(\dot{\alpha}) \Big|X\Big]\\
=& e(X)[(1-G(t_{k+1}))  + g(t_k) - (1-G(t_{k}))]  - e(X;\gamma^*) [(1-G(t_{k+1})) + g(t_k) -(1-G(t_{k}))]\\
=& e(X)\cdot 0 - e(X;\gamma^*)\cdot 0=0.
\end{align*}

If $\mathcal{M}_{\ps}$ is correctly specified, then $\gamma^*=\dot{\gamma}$ and $e(X)=e(X;\dot{\gamma})$. Hence,
\begin{align*}
&\mathbb{E}[(A-e(X))R_k(\alpha)|X] \\
=& \mathbb{E}\Big\{(A-e(X;\dot{\gamma}))\Big[ \frac{1}{S_0(t_k;{\alpha}^*,{\beta}^*)} H_k(\dot{\alpha}) + \frac{1}{S_0(t_k;{\alpha}^*,{\beta}^*)} C_k(\dot{\alpha})\\*
&\qquad {}- \frac{1}{S_0(t_{k-1};{\alpha}^*,{\beta}^*)} H_{k-1}(\dot{\alpha})\Big]\Big|X \Big \}\\
=&\frac{1}{S_0(t_k;{\alpha}^*,{\beta}^*)} \Big[ \mathbb{E}(AH_k(\dot{\alpha})|X) + \mathbb{E}(AC_k(\dot{\alpha})|X) \Big]\\*
&\qquad {}- \frac{1}{S_0(t_{k-1};{\alpha}^*,{\beta}^*)} \mathbb{E}(AH_{k-1}(\dot{\alpha})|X)\\
&-e(X;\dot{\gamma}) \mathbb{E}\Big[ \frac{1}{S_0(t_k;{\alpha}^*,{\beta}^*)} H_k(\dot{\alpha}) + \frac{1}{S_0(t_k;{\alpha}^*,{\beta}^*)} C_k(\dot{\alpha})\\*
&\qquad {}- \frac{1}{S_0(t_{k-1};{\alpha}^*,{\beta}^*)} H_{k-1}(\dot{\alpha}) \Big|X\Big]\\
=&\frac{1}{S_0(t_k;{\alpha}^*,{\beta}^*)}\Big[S_0(t_k) (1-G(t_{k+1})) e(X;\dot{\gamma}) + S_0(t_k) g(t_k) e(X;\dot{\gamma})\Big]\\*
&\qquad {}- \frac{1}{S_0(t_{k-1};{\alpha}^*,{\beta}^*)} S_0(t_{k-1}) (1-G(t_{k})) e(X;\dot{\gamma})\\
&- e(X;\dot{\gamma}) \Big[\frac{1}{S_0(t_k;{\alpha}^*,{\beta}^*)} (S_0(t_k) (1-G(t_{k+1})) + S_0(t_k) g(t_k))\\*
&\qquad {}- \frac{1}{S_0(t_{k-1};{\alpha}^*,{\beta}^*)} S_0(t_{k-1}) (1-G(t_{k})) \Big]\\
=& e(X;\dot{\gamma}) \Big[\frac{1}{S_0(t_k;{\alpha}^*,{\beta}^*)} S_0(t_k) (1-G(t_{k}))\\*
&\qquad {}- \frac{1}{S_0(t_k;{\alpha}^*,{\beta}^*)} S_0(t_k) (1-G(t_{k}))\Big] \\
&- e(X;\dot{\gamma}) \Big[\frac{1}{S_0(t_{k-1};{\alpha}^*,{\beta}^*)} S_0(t_{k-1}) (1-G(t_{k}))\\*
&\qquad {}- \frac{1}{S_0(t_{k-1};{\alpha}^*,{\beta}^*)} S_0(t_{k-1}) (1-G(t_{k}))\Big]=0.
\end{align*}
Therefore, $\mathbb{E}[U_{\sr}(\dot{\alpha};\alpha^*,\beta^*,\gamma^*;\mathcal{W})]=0$ under $\mathcal{M}_{\union}$.

Secondly, by applying Theorem 5.21 of \citet{van2000asymptotic}, we obtain the asymptotic normality of $\widehat{\alpha}_{\DR}(\mathcal{W})$ and the corresponding influence function is
	\begin{align}\label{supp-eq:if-sr}
	-\mathbb{E}\Big [\frac{\partial U_{\sr}(\alpha;\alpha^*,\beta^*,\gamma^*;\mathcal{W})}{\partial \alpha} \Big |_{\alpha=\dot{\alpha}}\Big]^{-1} \widetilde{U}_{\sr}(\dot{\alpha};\alpha^*,\beta^*,\gamma^*;\mathcal{W}),
\end{align}
where
\begin{align*}
	&\widetilde{U}_{\sr}(\dot{\alpha};\alpha^*,\beta^*,\gamma^*;\mathcal{W})
	= U_{\sr}(\dot{\alpha};\alpha^*,\beta^*,\gamma^*;\mathcal{W})\\
	&-\mathbb{E}\Big [\frac{\partial U_{\sr}(\dot{\alpha};\alpha^*,\beta^*,\gamma;\mathcal{W})}{\partial \gamma}\Big |_{\gamma=\gamma^*} \Big] \Big \{ \mathbb{E}\Big[ -X\frac{\partial e(X;\gamma)}{\partial \gamma} \Big |_{\gamma=\gamma^*} \Big] \Big \}^{-1} X \{A-e(X;\gamma^*)\}\\
	&- \mathbb{E}\Big [\frac{\partial U_{\sr}(\dot{\alpha};\alpha,\beta,\gamma^*;\mathcal{W})}{\partial (\alpha^\top, \beta^\top)^\top}\Big |_{
		\begin{subarray}{c}
			\alpha = \alpha^* \\
			\beta=\beta^*
		\end{subarray}} \Big ] \Big \{  \mathbb{E}\Big [\frac{\partial S_{\sr}(\alpha,\beta)}{\partial (\alpha^\top, \beta^\top)^\top} \Big |_{\begin{subarray}{c}
			\alpha = \alpha^* \\
			\beta=\beta^*
		\end{subarray}}\Big ]\Big\}^{-1} S_{\sr}(\alpha^*,\beta^*).
\end{align*}

Finally, when both $\mathcal{M}_{\sop}$ and
$\mathcal{M}_{\ps}$ are correctly specified, it holds true that $(\alpha^*,\beta^*)=(\dot{\alpha},\dot{\beta})$ and $\gamma^*=\dot{\gamma}$. Then, we have
\begin{align*}
&\mathbb{E}\Big [\frac{\partial U_{\sr}(\dot{\alpha};\alpha^*,\beta^*,\gamma;\mathcal{W}_{\eff})}{\partial \gamma}\Big |_{\gamma=\gamma^*} \Big] = \mathbb{E}\Big [\frac{\partial U_{\sr}(\dot{\alpha};\dot{\alpha},\dot{\beta},\gamma;\mathcal{W}_{\eff})}{\partial \gamma}\Big |_{\gamma=\dot{\gamma}} \Big]\\
=& \mathbb{E}\Big [ \mathcal{W}_{\text{eff}}(X)~ R(\dot{\alpha})(-\frac{\partial e(X;\gamma)}{\partial \gamma}) + (A-e(X;\gamma))(R(\dot{\alpha})^\top \otimes \mathbf{I}_{p_\alpha})\frac{\partial \text{Vec}(\mathcal{W}_{\text{eff}}(X))}{\partial \gamma} \Big |_{\gamma=\dot{\gamma}} \Big]\\
=&0,\\
&\mathbb{E}\Big [\frac{\partial U_{\sr}(\dot{\alpha};\alpha,\beta,\gamma^*;\mathcal{W}_{\eff})}{\partial (\alpha^\top, \beta^\top)^\top}\Big |_{
\begin{subarray}{c}
\alpha = \alpha^* \\
\beta=\beta^*
\end{subarray}} \Big ] = \mathbb{E}\Big [\frac{\partial U_{\sr}(\dot{\alpha};\alpha,\beta,\dot{\gamma};\mathcal{W}_{\eff})}{\partial (\alpha^\top, \beta^\top)^\top}\Big |_{
\begin{subarray}{c}
\alpha = \dot{\alpha} \\
\beta=\dot{\beta}
\end{subarray}} \Big ]\\
=& \mathbb{E}\Big [ (A-e(X;\dot\gamma))\mathcal{W}_{\eff}(X)\frac{\partial R(\dot{\alpha})}{\partial (\alpha^\top,\beta^\top)^\top} +(A-e(X;\dot\gamma))(R(\dot{\alpha})^\top \otimes \mathbf{I}_{p_\alpha})\frac{\partial \text{Vec}(\mathcal{W}_{\eff}(X))}{\partial (\alpha^\top,\beta^\top)^\top} \Big |_{
\begin{subarray}{c}
\alpha = \dot{\alpha} \\
\beta=\dot{\beta}
\end{subarray}} \Big ]\\
=&0.
\end{align*}
Therefore, the influence function of $\widehat{\alpha}_{\DR}(\mathcal{W}_{\eff})$ is $$-\mathbb{E}\Big [\frac{\partial U_{\sr}(\alpha;\dot\alpha,\dot\beta,\dot\gamma;\mathcal{W}_{\eff})}{\partial \alpha} \Big |_{\alpha=\dot{\alpha}}\Big]^{-1} {U}_{\sr}(\dot{\alpha};\dot\alpha,\dot\beta,\dot\gamma;\mathcal{W}_{\eff}).$$

Following Remark 3.2 of \citet{richardson2017modeling} and the argument of \citet{robins1994estimation}, \(S^*(\dot\alpha)\) remains the efficient score under \(\mathcal{M}_{\union}\) when \(\mathcal{M}_{\ps}\) is correctly specified. Since
\[
U_{\sr}(\dot{\alpha};\dot{\alpha},\dot{\beta},\dot{\gamma};\mathcal{W}_{\eff})=S^*(\dot{\alpha}),
\]
the influence function above is the corresponding efficient influence function under \(\mathcal{M}_{\union}\). Therefore, if both $\mathcal{M}_{\sop}$ and
$\mathcal{M}_{\ps}$ are correctly specified,  \(\widehat{\alpha}_{\DR}(\mathcal{W}_{\eff})\) attains the semiparametric efficiency bound under \(\mathcal{M}_{\union}\).

\end{proof}

The covariance matrix of \(\widehat{\alpha}_{\DR}(\mathcal W)\) is estimated using the influence function in \eqref{supp-eq:if-sr}, and the Wald confidence intervals are constructed accordingly.

\subsection[Data application: additional results for the prostate cancer trial application]{Data application: additional results for the\\prostate cancer trial application}\label{supp-sec:prostate-additional}

\begin{figure}[!htbp]
	\centering
	\includegraphics[width=0.80\textwidth]
	{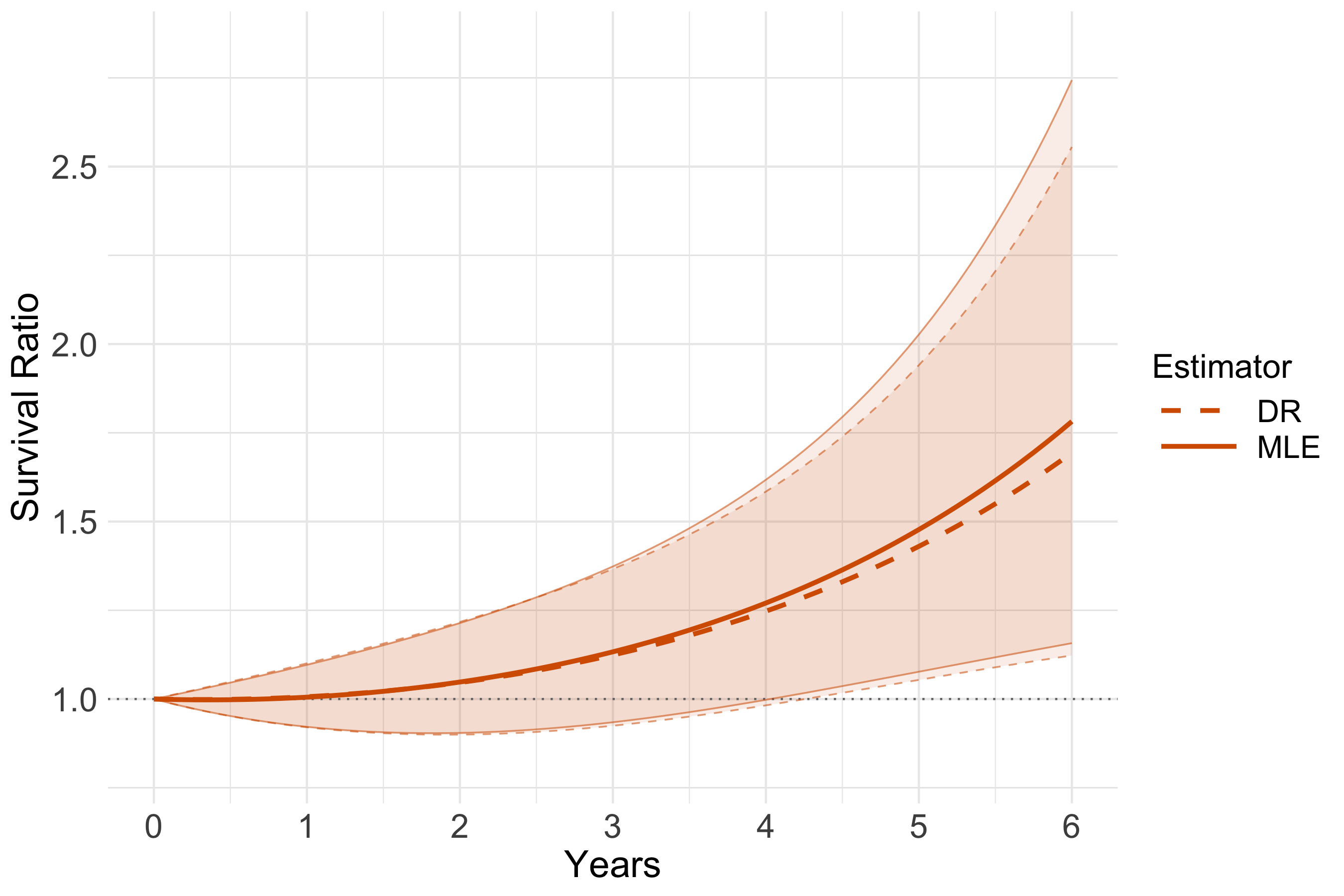}
	\caption{Estimated survival-ratio curves comparing 1.0 mg estrogen
		with placebo. Solid and dashed curves denote the maximum likelihood and doubly robust estimates, respectively; the corresponding shaded regions show pointwise
		95\% confidence intervals.}
	\label{supp-fig:prostate-1mg-ci}
	\par\small\noindent\textit{Alt text: Maximum likelihood and doubly
		robust fitted survival-ratio curves comparing 1.0 mg estrogen with
		placebo over six years. Both curves increase during follow-up, and
		the pointwise confidence intervals widen over time.}
\end{figure}

\begin{figure}[!htbp]
	\centering
	\includegraphics[width=0.95\textwidth]
	{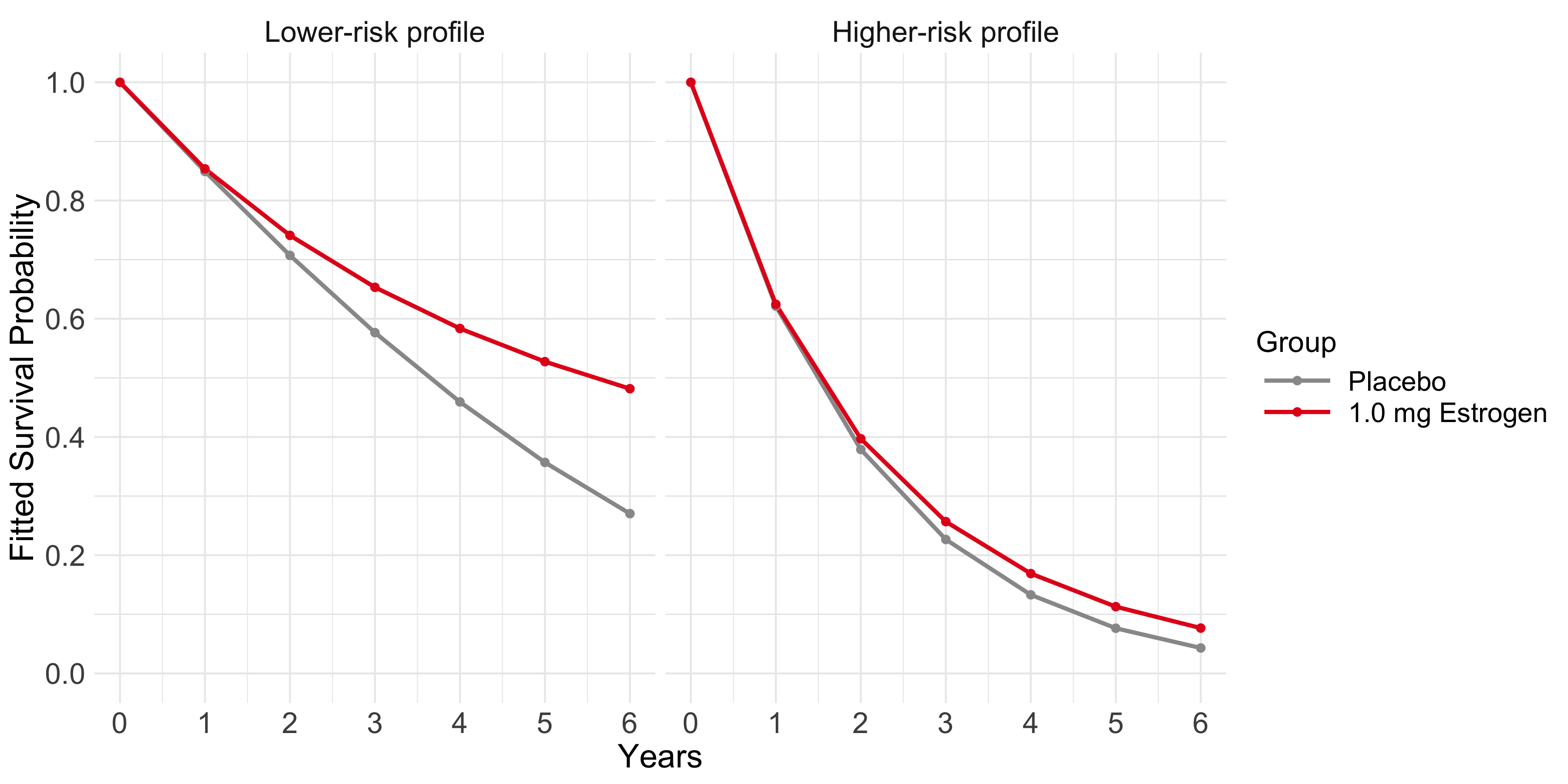}
	\caption{Maximum likelihood fitted survival curves under placebo and
		1.0 mg estrogen for two illustrative covariate profiles.}
	\label{supp-fig:prostate-profile-survival}
	\par\small\noindent\textit{Alt text: Fitted survival curves under
		placebo and 1.0 mg estrogen are shown for lower-risk and
		higher-risk covariate profiles. Fitted survival is substantially
		lower for the higher-risk profile, while the 1.0 mg estrogen curve
		lies above the placebo curve in both profiles during later
		follow-up.}
\end{figure}

Although the fitted survival ratio does not vary with \(X\), the
treatment-specific survival functions remain covariate-dependent.
To illustrate this distinction, we calculated the maximum likelihood
fitted survival curves under two covariate profiles. The lower-risk
profile represents a 70-year-old patient with normal daily activity
and no history of cardiovascular disease, whereas the higher-risk
profile represents a 76-year-old patient without normal daily activity
and with a history of cardiovascular disease. Serum hemoglobin was
fixed at its median value of $13.7$ for both profiles. Figure~\ref{supp-fig:prostate-profile-survival} shows that, compared with the lower-risk profile, the higher-risk profile has substantially lower fitted survival under both placebo and 1.0 mg estrogen. Within each profile, the treatment-specific curves are similar at the beginning of follow-up but gradually separate, with higher fitted survival under 1.0 mg estrogen than under placebo.

\subsection{Semiparametric estimation under \texorpdfstring{$C\ind T \mid (A,X)$}{C independent of T given (A,X)}}\label{supp-sec:sr-censoring}
In this part, we assume $C\ind T \mid (A,X)$ and let $g_a(t;X)=\pr(C=t \mid X,A=a)$ and $G_a(t_k;X)=\sum_{i=1}^{k-1}g_a(t_i;X)$.

\begin{proposition}\label{supp-prop:effScore}
Under the observed-data law restricted by the model $\mathcal{M}_{\sr}$, the efficient score for $\alpha$ is
\begin{align*}
\widetilde{S}^*(\alpha) = \widetilde{\mathcal{W}}_{\text{eff}}(X)~ \widetilde{R}(\alpha)(A-e(X)),
\end{align*}
where $\widetilde{\mathcal{W}}_{\eff}(X)$ is the efficient weighting $p_\alpha\times m$ matrix,
\begin{align*}
\widetilde{R}(\alpha) &= V_{A, X}^{-1}R(\alpha),\quad\text{and}\\
V_{A,X} &= \text{diag}(v_{A,X}^1,...,v_{A,X}^m)
\end{align*}
with $v_{A,X}^k = 1-G_A(t_k;X)$.
\end{proposition}

\begin{proof}[Proof of Proposition \ref{supp-prop:effScore}]
Replacing $G(t)$ with $G_A(t)$ and following Steps 1--5 in the proof of Theorem \ref{thm:sur:effscore}, the efficient score for $\alpha$ becomes
\begin{align*}
\widetilde{S}^*(\alpha) = \mathcal{F}_{\sr,X}^\top \Gamma \Big (A \mathbf{I}_m   - \mathbb{E}(A  Q_{A,X} V_{A,X} \mid X) \Big [ \mathbb{E}(Q_{A,X}V_{A,X}  \mid X) \Big ]^{-1} \Big )Q_{A,X} R(\alpha).
\end{align*}
Similarly, we can prove that
\begin{align*}
&\Big (A \mathbf{I}_m   - \mathbb{E}(A  Q_{A,X} V_{A,X} \mid X) \Big [ \mathbb{E}(Q_{A,X}V_{A,X}  \mid X) \Big ]^{-1} \Big )Q_{A,X}V_{A,X}\\
=& (A-e(X)) \widetilde{D}_X \mathbb{E}(A  Q_{A,X} V_{A,X} \mid X) \Big [\mathbb{E}(  Q_{A,X} V_{A,X} \mid X) \Big ]^{-1},
\end{align*}
which implies that
\begin{align*}
\widetilde{S}^*(\alpha) =& \mathcal{F}_{\sr,X}^\top \Gamma \widetilde{D}_X \mathbb{E}(A  \widetilde{Q}_{A,X}  \mid X) \Big [\mathbb{E}(  \widetilde{Q}_{A,X} \mid X) \Big ]^{-1} \widetilde{R}(\alpha)(A-e(X))\\
=& \widetilde{\mathcal{W}}_{\eff}(X)\widetilde{R}(\alpha)(A-e(X)),
\end{align*}
where $\widetilde{\mathcal{W}}_{\eff}(X)=\mathcal{F}_{\sr,X}^\top \Gamma \widetilde{D}_X \mathbb{E}(A  \widetilde{Q}_{A,X}  \mid X) \Big [\mathbb{E}(  \widetilde{Q}_{A,X} \mid X) \Big ]^{-1}$, $\widetilde{Q}_{A,X} = Q_{A,X} V_{A,X}$ and $\widetilde{D}_X=\text{diag}(\tilde{d}_X^1,...,\tilde{d}_X^m)$ with $\tilde{d}_X^k= \frac{S_0(t_k)}{e(X)(1-\frac{S_0(t_k)}{S_0(t_{k-1})})}(1-G_0(t_k))$.
\end{proof}

\FloatBarrier
\section{Cumulative incidence ratio modeling}\label{supp-sec:cir}

\subsection{Properties of the backward odds product}\label{supp-sec:cir-model}
\begin{remark}\label{supp-remark:cir}(Choice of $\Lambda(t;X,\rho)$ and $\Xi(t_{k+1}, t_k; X, \tau)$) The constraints for the cumulative distribution function $F_a(t;X)$ imply the following conditions for $\CIR(t;X)$ and $\widetilde{\SOP}(t_{k+1},t_k;X)$:
	\[
	\CIR(t;X)\rightarrow 1 \quad \text{as } t\rightarrow+\infty;
	\]
for fixed $t_{k+1}$ such that \(F_a(t_{k+1};X)>0\), \(a=0,1\),
\[
\widetilde{\SOP}(t_{k+1},t_k;X)\rightarrow 0
\quad \text{as } t_k\rightarrow 0^+.
\]

For example, our choice for $f_{\cir}(t;X)$ can be $f_{\cir}(t;X)=\frac{1}{t}\cdot X$ or $f_{\cir}(t;X)=\exp(-t) \cdot X$. In addition, we let $\Xi(t_{k+1},t_k;X,\tau)=\tau^\top X - \log({1}/{t_k} - {1}/{t_{k+1}})$.

\end{remark}

\subsection{Maximum likelihood estimation}\label{supp-sec:cir-mle}
Assume $C \ind (A,T)|X$ and then the log-likelihood function for an individual is
\begin{align*}
l_{\cir}(\rho,\tau) = \sum_{k=1}^{m}  \Big\{ &{\mathbb{I}(Y = t_{k}) \Delta} \log \Big  ( (
 F_{A}(t_{k}; X) - F_{A}(t_{k-1}; X) ) (1 - G(t_{k}; X)) \Big) \\
 &+ {\mathbb{I}(Y = t_{k})(1-\Delta)} \log \Big ( (1-F_{A}(t_{k}; X)) g(t_{k}; X) \Big )  \Big\}.
\end{align*}
Let the maximum likelihood estimators $\widehat{\rho}$ and $\widehat{\tau}$ solve
\begin{align*}
	0=\mathbb{P}_n  [ S_{\cir}(\rho,\tau)],
\end{align*}
where $S_{\cir}(\rho,\tau)=\frac{\partial}{\partial (\rho^\top,\tau^\top)^\top} l_{\cir}(\rho,\tau)$ is the score function for $\rho$ and $\tau$, and $\mathbb{P}_n$ denotes the empirical average. As in \ref{supp-sec:sr-mle}, inference for $\rho$ and $\tau$ is based on a sandwich variance estimator \citep{white1982maximum}.

\subsection{Doubly robust estimation}\label{supp-sec:cir-dr}

\begin{proof}[Proof of Theorem \ref{thm:sur:effscore-CIR}]
Consider
\begin{align*}
&\log P(Y,\Delta,A,X)\\
=&  \sum_{k=1}^m \prod_{j=0}^{k-1}(1- \mathbb{I}(Y=t_j))\Big [ \mathbb{I}(Y=t_k)\Delta \log \Big( 1- \frac{1-F_A(t_k)}{1-F_A(t_{k-1})} \Big)\\*
&\qquad {}+  (1-\mathbb{I}(Y=t_k)\Delta)\log \Big( \frac{1-F_A(t_k)}{1-F_A(t_{k-1})} \Big)\Big]\\
&+ \sum_{k=1}^m \Big \{ \mathbb{I}(Y=t_k) \Delta \log \Big ( 1- G(t_k) \Big ) + \mathbb{I}(Y=t_k) (1-\Delta) \log(  g(t_k)  )\Big \}+ \log P(A,X),
\end{align*}
and under model $\mathcal{M}_{\cir}$, we compute the score function
\begin{align*}
&S_{\rho} = \frac{\partial \log P(Y,\Delta,A,X)}{\partial \rho}\\
=&\sum_{k=1}^m \prod_{j=0}^{k-1}(1- \mathbb{I}(Y=t_j))\Big [ \mathbb{I}(Y=t_k)\Delta  \frac{-1}{ 1- \frac{1-F_A(t_k)}{1-F_A(t_{k-1})}} \frac{\partial (\frac{1-F_A(t_k)}{1-F_A(t_{k-1})})}{\partial \rho} \\
&+  (1-\mathbb{I}(Y=t_k)\Delta)\frac{1}{\frac{1-F_A(t_k)}{1-F_A(t_{k-1})}} \frac{\partial (\frac{1-F_A(t_k)}{1-F_A(t_{k-1})})}{\partial \rho} \Big]\\
=&\sum_{k=1}^m \frac{\partial (\frac{1-F_A(t_k)}{1-F_A(t_{k-1})})}{\partial \rho} \prod_{j=0}^{k-1}(1- \mathbb{I}(Y=t_j)) \frac{1-\mathbb{I}(Y=t_k)\Delta - \frac{1-F_A(t_k)}{1-F_A(t_{k-1})}}{(1- \frac{1-F_A(t_k)}{1-F_A(t_{k-1})}) \frac{1-F_A(t_k)}{1-F_A(t_{k-1})}}\\
=&A\sum_{k=1}^m \frac{-F_A(t_k)f_{\cir}(t_k)(1-F_A(t_{k-1}))+F_A(t_{k-1})f_{\cir}(t_{k-1})(1-F_A(t_k))}{(1-F_A(t_{k-1}))^2}\\*
&\qquad \frac{1-F_A(t_k)}{(1- \frac{1-F_A(t_k)}{1-F_A(t_{k-1})}) \frac{1-F_A(t_k)}{1-F_A(t_{k-1})}}\\*
& \times\Big[ \frac{1}{1-F_A(t_k)}\prod_{j=0}^k (1-\mathbb{I}({Y=t_j})) + \frac{1}{1-F_A(t_k)} (1-\Delta)\mathbb{I}(Y=t_k)\\*
&\qquad {}- \frac{1}{1-F_A(t_{k-1})}\prod_{j=0}^{k-1} (1-\mathbb{I}({Y=t_j})) \Big]\\
=& A\sum_{k=1}^m \frac{-F_A(t_k)(1-F_A(t_{k-1}))f_{\cir}(t_k)+F_A(t_{k-1})(1-F_A(t_k))f_{\cir}(t_{k-1})}{F_A(t_k)-F_A(t_{k-1})}\\*
& \times\Big[ \frac{1}{1-F_A(t_k)}\prod_{j=0}^k (1-\mathbb{I}({Y=t_j})) + \frac{1}{1-F_A(t_k)} (1-\Delta)\mathbb{I}(Y=t_k)\\*
&\qquad {}- \frac{1}{1-F_A(t_{k-1})}\prod_{j=0}^{k-1} (1-\mathbb{I}({Y=t_j})) \Big]\\
=& A \mathcal{F}_{\cir,X}^\top Q_{A,X} R(\rho),
\end{align*}
where $\rho\in\mathbb{R}^{p_\rho}$, $\mathcal{F}_{\cir,X} = (f_{\cir}(t_1),...,f_{\cir}(t_m))^\top \in \mathbb{R}^{m\times p_\rho}$, $Q_{A,X}=Q_{A,X}^1 + Q_{A,X}^2 \Gamma $, $Q_{A,X}^1=\text{diag}(q_1^1,..,q_m^1)$, $Q_{A,X}^2=\text{diag}(q_1^2,..,q_m^2)$, and $\Gamma = (\Gamma_{ij})_{m \times m}$ with
\begin{align*}
&q_i^1 = -\frac{F_A(t_i)(1-F_A(t_{i-1}))}{F_A(t_i)-F_A(t_{i-1})}, \\
& q_i^2 = \frac{F_A(t_i)(1-F_A(t_{i+1}))}{F_A(t_{i+1})-F_A(t_i)},\\
&\Gamma_{ij}=
\begin{cases}
1 & \text{ if }  j=i+1\\
0 & \text{ otherwise},
\end{cases}
\end{align*}
and $R(\rho)=(R_1(\rho),\cdots, R_m(\rho))^\top$ with
\begin{align*}
R_k(\rho) ={}& \frac{1}{1-F_A(t_k)}\prod_{j=0}^k (1-\mathbb{I}({Y=t_j})) + \frac{1}{1-F_A(t_k)} (1-\Delta)\mathbb{I}(Y=t_k)\\*
&{}- \frac{1}{1-F_A(t_{k-1})}\prod_{j=0}^{k-1} (1-\mathbb{I}({Y=t_j})).
\end{align*}

Similar to the steps in the proof of Theorem \ref{thm:sur:effscore}, we consider an arbitrary parametric submodel by substituting $F_0(t;\zeta_1)$, $g(t;\zeta_2)$ and $P(A,X; \zeta_3)$ into $P(Y,\Delta,A,X)$, where $\zeta_1$, $\zeta_2$ and $\zeta_3$ are finite-dimensional nuisance parameters. Then, the nuisance tangent space $\Lambda$ can be written as a direct sum of three orthogonal spaces $\Lambda=\Lambda_{1s}\oplus \Lambda_{2s}\oplus \Lambda_{3s}$, where
\[
\Lambda_{js}=\text{mean-square closure of }\{BS_{\zeta_j}(Y,\Delta,A,X)\text{ for all }B\},
\]
and
\[
S_{\zeta_j}(Y,\Delta,A,X)=\partial \log P(Y,\Delta,A,X)/\partial \zeta_j
\]
for $j=1,2,3$. On the other hand, the tangent space $\mathcal{H}$ obtained by leaving $F_A(t)$ unrestricted can be written as a direct sum of three orthogonal spaces $\mathcal{H}=\Lambda_{1s}^*\oplus \Lambda_{2s}\oplus \Lambda_{3s}$, where $\Lambda_{1s}^*$ is the space associated with $F_A(t)$. Here, $\Lambda^\perp$ is taken within $\mathcal{H}$.

To derive the space $\Lambda_{1s}$, we consider the parametric submodel $\log F_0(t;\zeta_1)=\log F_0(t)+\zeta_1^\top h(t)$, where $h(t)$ is an arbitrary function. Then, we obtain the score function with respect to $\zeta_1$ $S_{\zeta_1}= \mathcal{H}_X^\top Q_{A,X} R(\rho)$, where $\mathcal{H}_X=(h(t_1)^\top,...,h(t_m)^\top)^\top$, and therefore we have $\Lambda_{1s}=\Big\{\mathcal{H}_X^\top Q_{A,X} R(\rho): \mathcal{H}_X \Big\}$. Similarly, to derive the space $\Lambda_{1s}^*$, we consider the parametric submodel $\log F_A(t;\zeta_1^*)=\log F_A(t)+(\zeta_1^*)^\top b_A(t)$, where $b_A(t)$ is an arbitrary function, and compute $S_{\zeta_1^*}=\mathcal{B}_{A,X}^\top Q_{A,X} R(\rho)$, where $\mathcal{B}_{A,X}=(b_A(t_1)^\top,...,b_A(t_m)^\top)^\top$. Thus, $\Lambda_{1s}^*=\Big\{\mathcal{B}_{A,X}^\top Q_{A,X} R(\rho): \mathcal{B}_{A,X}\Big\}$.

Next, we compute $\Lambda^\perp$ by finding $\mathcal{H}_X^*$ such that
\[
\mathbb{E}\Big[ (\mathcal{B}_{A,X}^\top J_{A,X}(\rho)-(\mathcal{H}^*_X)^\top J_{A,X}(\rho) )^\top \mathcal{H}_X^\top J_{A,X}(\rho)\Big]=0
\]
for all $\mathcal{H}_X$, where $J_{A,X}(\rho)=Q_{A,X}R(\rho)$. Similar to Step 5 in the proof of Theorem \ref{thm:sur:effscore}, we obtain $\mathcal{H}_X^* = \Big[\mathbb{E}(\Sigma_{A,X}|X)\Big]^{-1}\mathbb{E}(\Sigma_{A,X}\mathcal{B}_{A,X}|X)$, where
\begin{align*}
\Sigma_{A,X}=&\mathbb{E}(J_{A,X}(\rho)(J_{A,X}(\rho))^\top |A,X)\\
=& Q_{A,X}\widetilde{\Sigma}_{A,X} Q_{A,X}^\top,
\end{align*}
where $\widetilde{\Sigma}_{A,X}$ is a diagonal matrix with diagonal element $\widetilde{\sigma}^k_{A,X}=(1-G(t_k))(\frac{1}{1-F_A(t_k)}-\frac{1}{1-F_A(t_{k-1})})$, and hence $\Lambda^\perp = \Big\{(\mathcal{B}_{A,X}-\mathcal{H}_X^*)^\top J_{A,X}(\rho): \mathcal{B}_{A,X}\Big\}$. Therefore, the efficient score for $\rho$ is
\begin{align}\label{supp-eq:cir:eff1}
 S^*(\rho)=&\mathcal{F}_{\cir,X}^\top \Big(A \mathbf{I}_m - \mathbb{E}(A\Sigma_{A,X}|X)\Big[\mathbb{E}(\Sigma_{A,X}|X)\Big]^{-1}\Big) J_{A,X}(\rho)\nonumber\\
 =&\mathcal{F}_{\cir,X}^\top \Big(A \mathbf{I}_m - e(X)\Sigma_{1,X}\widebar{\Sigma}_{X}^{-1}\Big) J_{A,X}(\rho),
 \end{align}
where $\widebar{\Sigma}_{X}=e(X)\Sigma_{1,X} + (1-e(X))\Sigma_{0,X}$. 

Finally, we show that
\begin{align}\label{supp-eq:cir:eff2}
S^*(\rho)=\mathcal{F}_{\cir,X}^\top \Omega_X \left[ \frac{A}{e(X)}\Sigma_{1,X}^{-1} J_{1,X}(\rho) - \frac{1-A}{1-e(X)}\Sigma_{0,X}^{-1} J_{0,X}(\rho)  \right],
\end{align}
where $\Omega_X = e(X)(1-e(X)) \Sigma_{1,X}\widebar{\Sigma}_{X}^{-1}\Sigma_{0,X}$. When $A=1$, it follows from the right hand side of \eqref{supp-eq:cir:eff2} that
\begin{align*}
&\mathcal{F}_{\cir,X}^\top \Omega_X \left[ \frac{1}{e(X)}\Sigma_{1,X}^{-1} J_{1,X}(\rho) \right]\\
=&\mathcal{F}_{\cir,X}^\top (e(X)(1-e(X)) \Sigma_{1,X}\widebar{\Sigma}_{X}^{-1}\Sigma_{0,X}) \left[ \frac{1}{e(X)}\Sigma_{1,X}^{-1} J_{1,X}(\rho) \right]\\
=& (1-e(X))\mathcal{F}_{\cir,X}^\top \Sigma_{1,X}\widebar{\Sigma}_{X}^{-1}\Sigma_{0,X} \Sigma_{1,X}^{-1} J_{1,X}(\rho),
\end{align*}
and it follows from the right hand side of \eqref{supp-eq:cir:eff1} that
$
\mathcal{F}_{\cir,X}^\top \Big( \mathbf{I}_m - e(X)\Sigma_{1,X}\widebar{\Sigma}_{X}^{-1}\Big) J_{1,X}(\rho).
$
Thus, it suffices to show $(1-e(X))\Sigma_{1,X}\widebar{\Sigma}_{X}^{-1}\Sigma_{0,X} \Sigma_{1,X}^{-1} = \mathbf{I}_m - e(X)\Sigma_{1,X}\widebar{\Sigma}_{X}^{-1}$, which is equivalent to showing $(1-e(X))\Sigma_{1,X}\widebar{\Sigma}_{X}^{-1}\Sigma_{0,X}  = \Sigma_{1,X} - e(X)\Sigma_{1,X}\widebar{\Sigma}_{X}^{-1}\Sigma_{1,X}$. Since $(1-e(X))\Sigma_{0,X}=\widebar{\Sigma}_X-e(X)\Sigma_{1,X}$, we have
\begin{align*} &(1-e(X))\Sigma_{1,X}\widebar{\Sigma}_{X}^{-1}\Sigma_{0,X}\\
=&\Sigma_{1,X}\widebar{\Sigma}_{X}^{-1}[\widebar{\Sigma}_X-e(X)\Sigma_{1,X}]\\
=&\Sigma_{1,X} - e(X)\Sigma_{1,X}\widebar{\Sigma}_{X}^{-1}\Sigma_{1,X}.
\end{align*}
When $A=0$, it follows from the right-hand side of \eqref{supp-eq:cir:eff2} that
\begin{align*}
&\mathcal{F}_{\cir,X}^\top \Omega_X \left[- \frac{1}{1-e(X)}\Sigma_{0,X}^{-1} J_{0,X}(\rho)  \right]\\
=&\mathcal{F}_{\cir,X}^\top (e(X)(1-e(X)) \Sigma_{1,X}\widebar{\Sigma}_{X}^{-1}\Sigma_{0,X}) \left[- \frac{1}{1-e(X)}\Sigma_{0,X}^{-1} J_{0,X}(\rho)  \right]\\
=&-e(X)\mathcal{F}_{\cir,X}^\top \Sigma_{1,X}\widebar{\Sigma}_{X}^{-1} J_{0,X}(\rho),
\end{align*}
whereas the right-hand side of \eqref{supp-eq:cir:eff1} is $\mathcal{F}_{\cir,X}^\top \big\{-e(X)\Sigma_{1,X}\widebar{\Sigma}_{X}^{-1}\big\}J_{0,X}(\rho)$. Therefore, \eqref{supp-eq:cir:eff2} holds.

With $\widetilde{\mathcal{W}}_{\eff}(X)=\mathcal{F}_{\cir,X}^\top \Omega_X$ and $Z_a(\rho)=\Sigma_{a,X}^{-1} J_{a,X}(\rho)$ for $a=0,1$, we have
\begin{align*}
S^*(\rho) = \widetilde{\mathcal{W}}_{\eff}(X) \left[ \frac{A}{e(X)}Z_1(\rho) - \frac{1-A}{1-e(X)}Z_0(\rho)  \right].
\end{align*}
This completes the proof.
\end{proof}

Because the efficient score for $\rho$ depends on the censoring distribution, we introduce
\[
\mathcal{M}_{\cen}:\quad G(t;X)=G(t;X,\eta),
\]
where $\eta$ is an unknown finite-dimensional parameter. We use a discrete-time logistic cen\-sor\-ing-hazard model. Let $l_{\cen}(\eta)$ and $S_{\cen}(\eta)=\partial l_{\cen}(\eta)/\partial\eta$ denote its log-likelihood and score, and let $\widehat\eta$ solve $\mathbb{P}_n\{S_{\cen}(\eta)\}=0$.

Let $\mathcal W(X)$ be a weighting matrix for which the Jacobian of the expected estimating function with respect to $\rho$ is nonsingular at the true value. Motivated by Theorem \ref{thm:sur:effscore-CIR}, we define $\widehat\rho_{\DR}(\mathcal W)$ as a solution to
\begin{align*}
	0=\mathbb{P}_n[U_{\cir}(\rho;\widehat{\rho},\widehat{\tau},\widehat{\gamma},\widehat{\eta};\mathcal{W})],
\end{align*}
where $$U_{\cir}(\rho;\widehat{\rho},\widehat{\tau},\widehat{\gamma},\widehat{\eta};\mathcal{W})=\mathcal{W}(X) \left[\frac{A}{e(X;\widehat{\gamma})}Z_1(\rho;\widehat{\rho},\widehat{\tau},\widehat{\eta})-\frac{1-A}{1-e(X;\widehat{\gamma})}Z_0(\rho;\widehat{\rho},\widehat{\tau},\widehat{\eta})\right],$$ $\mathcal{W}(X)$ is a weighting matrix, and $Z_a(\rho;\widehat{\rho},\widehat{\tau},\widehat{\eta})$ denotes the plug-in version of $Z_a$, obtained by replacing $F_a(t_k;X)$ with
\[
F_a(t_k;X,\rho,\widehat{\rho},\widehat{\tau})
=
\exp\{a \rho^\top f_{\cir}(t_k;X)\}
F_0(t_k;X,\widehat{\rho},\widehat{\tau})
\]
and replacing $G(t_k;X)$ with $G(t_k;X, \widehat{\eta})$, for $k=1,\ldots,m$.

\begin{remark}(Efficient weighting matrix $\widetilde{\mathcal W}_{\eff}(X; \widehat{\rho}, \widehat{\tau},\widehat\gamma, \widehat{\eta})$) Let $\widetilde{\mathcal W}_{\eff}(X; \widehat{\rho}, \widehat{\tau},\widehat{\gamma},\widehat{\eta})$ denote the plug-in version of $\widetilde{\mathcal W}_{\eff}(X)$, obtained by replacing $F_a(t_k;X)$ with $F_a(t_k;X,\widehat\rho,
\widehat\tau)$, replacing $e(X)$ with $e(X;\widehat\gamma)$, and replacing $G(t_k;X)$ with $G(t_k;X, \widehat{\eta})$, for $k=1,\ldots,m$ and $a=0,1$. Then, by setting $\mathcal{W}(X)=\widetilde{\mathcal W}_{\eff}(X; \widehat{\rho}, \widehat{\tau}, \widehat\gamma, \widehat{\eta})$, we obtain the estimator $\widehat{\rho}_{\DR}(\widetilde{\mathcal W}_{\eff})$.
\end{remark}

\begin{proof}[Proof of Proposition \ref{prop:sur:DR-CIR}]

It follows from Slutsky's theorem that the probability limit of $U_{\cir}(\rho;\allowbreak\widehat{\rho}, \allowbreak\widehat{\tau}, \allowbreak\widehat{\gamma}, \allowbreak\widehat{\eta}; \allowbreak\mathcal{W})$ is
\begin{align*}
	U_{\cir}(\rho;{\rho}^*, {\tau}^*, {\gamma}^*, {\eta}^*; \mathcal{W}):=\mathcal{W}(X)\left[ \frac{A}{e(X;\gamma^*)}Z_1(\rho;{\rho}^*,\tau^*,\eta^*) - \frac{1-A}{1-e(X;\gamma^*)}Z_0(\rho;{\rho}^*,\tau^*,\eta^*)  \right],
\end{align*}
where $\rho^*,\tau^*,\eta^*$ and $\gamma^*$ are the probability limits of $\widehat{\rho}, \widehat{\tau}, \widehat{\eta}$ and $\widehat{\gamma}$, respectively. Let $\dot{\rho},\dot{\tau},\dot{\eta}$ and $\dot{\gamma}$ be the unknown true values of $\rho,\tau,\eta$ and $\gamma$, respectively. When $\widetilde{\mathcal M}_{\sop}$ is correctly specified, $(\rho^*,\tau^*)=(\dot{\rho},\dot{\tau})$; when ${\mathcal M}_{\ps}$ is correctly specified, $\gamma^*=\dot{\gamma}$; when $\mathcal{M}_{\cen}$ is correctly specified, $\eta^*=\dot{\eta}$. We then show that $\mathbb{E}[U_{\cir}(\dot{\rho};\rho^*,\tau^*,\gamma^*,{\eta}^*)]=0$ under $\widetilde{\mathcal{M}}_{\union}$.

Note that
\begin{align}\label{supp-eq:residual}
&\mathbb{E}[R_k(\dot{\rho};\rho^*,\tau^*)|A=a,X]\nonumber\\
=&\frac{1}{1-F_a(t_k;\dot{\rho},\rho^*,\tau^*)}\pr(T>t_k,C\geq t_k | A=a,X)\nonumber\\
&-\frac{1}{1-F_a(t_{k-1};\dot{\rho},\rho^*,\tau^*)}\pr(T>t_{k-1},C\geq t_k | A=a,X)\nonumber\\
=&\frac{(1-F_a(t_k))(1-G(t_k))}{1-F_a(t_k;\dot{\rho},\rho^*,\tau^*)}  -\frac{(1-F_a(t_{k-1}))(1-G(t_k))}{1-F_a(t_{k-1};\dot{\rho},\rho^*,\tau^*)}\nonumber\\
=& (1-G(t_k))\left[\frac{1-F_a(t_k)}{1-F_a(t_k;\dot{\rho},\rho^*,\tau^*)}-\frac{1-F_a(t_{k-1})}{1-F_a(t_{k-1};\dot{\rho},\rho^*,\tau^*)}\right].
\end{align}

If $\widetilde{\mathcal{M}}_{\sop}$ is correctly specified, $(\rho^*,\tau^*)=(\dot{\rho},\dot{\tau})$ and $F_A(t)=F_A(t;\dot{\rho},\dot{\rho},\dot{\tau})$. Then, we have
\begin{align*}
	\mathbb{E}[R_k(\dot{\rho};\rho^*,\tau^*)|A=a,X]=0,
\end{align*}
which implies that
\begin{align*}
\mathbb{E}[Z_a(\dot{\rho};\rho^*,\tau^*,\eta^*)|A=a,X]=0.
\end{align*}
Thus,
\begin{align*}
&\mathbb{E} \left[ \frac{A}{e(X;\gamma^*)}Z_1(\dot\rho;\rho^*,\tau^*,\eta^*) - \frac{1-A}{1-e(X;\gamma^*)}Z_0(\dot\rho;\rho^*,\tau^*,\eta^*) \Big| X  \right]\\
=& \frac{e(X)}{e(X;\gamma^*)} \mathbb{E}[Z_1(\dot{\rho};\rho^*,\tau^*,\eta^*)|A=1,X] - \frac{1-e(X)}{1-e(X;\gamma^*)} \mathbb{E}[Z_0(\dot{\rho};\rho^*,\tau^*,\eta^*)|A=0,X] =0.
\end{align*}

If $\mathcal{M}_{\ps}\cap \mathcal{M}_{\cen}$ is correctly specified, then $(\gamma^*,\eta^*)=(\dot{\gamma},\dot{\eta})$, $e(X)=e(X;\dot{\gamma})$ and $g(t)=g(t;\dot{\eta})$. Under $\mathcal{M}_{\cir}$, we have
\begin{align*}
	\frac{F_1(t_k)}{F_0(t_k)}=\exp(\dot{\rho}^\top f_{\cir}(t_k)),
\end{align*}
and it holds true that
\begin{align*}
	\frac{F_1(t_k;\dot{\rho},\rho^*,\tau^*)}{F_0(t_k;\dot{\rho},\rho^*,\tau^*)}=\exp(\dot{\rho}^\top f_{\cir}(t_k)).
\end{align*}
Then, let $d_X=(d_{1,X},...,d_{m,X})^\top$ with $d_{k,X}=F_0(t_k)/F_0(t_k;\dot{\rho},\rho^*,\tau^*)-1$, and we have
\begin{align*}
1-F_a(t_k)=1-F_a(t_k;\dot{\rho},\rho^*,\tau^*)(1+d_{k,X}).
\end{align*}
It follows from \eqref{supp-eq:residual} that
\begin{align*}
&\mathbb{E}[R_k(\dot{\rho};\rho^*,\tau^*)|A=a,X]\\
=& (1-G(t_k))\left[\frac{1-F_a(t_k)}{1-F_a(t_k;\dot{\rho},\rho^*,\tau^*)}-\frac{1-F_a(t_{k-1})}{1-F_a(t_{k-1};\dot{\rho},\rho^*,\tau^*)}\right]\\
=&(1-G(t_k))\frac{\begin{aligned}
&(1-F_a(t_k))(1-F_a(t_{k-1};\dot{\rho},\rho^*,\tau^*))\\
&{}-(1-F_a(t_{k-1}))(1-F_a(t_k;\dot{\rho},\rho^*,\tau^*))
\end{aligned}}{(1-F_a(t_k;\dot{\rho},\rho^*,\tau^*))(1-F_a(t_{k-1};\dot{\rho},\rho^*,\tau^*))}\\
=&(1-G(t_k))\\*
&\times \frac{\begin{aligned}
&-F_a(t_k;\dot{\rho},\rho^*,\tau^*)(1-F_a(t_{k-1};\dot{\rho},\rho^*,\tau^*))d_{k,X}\\
&{}+F_a(t_{k-1};\dot{\rho},\rho^*,\tau^*)(1-F_a(t_{k};\dot{\rho},\rho^*,\tau^*))d_{k-1,X}
\end{aligned}}{(1-F_a(t_k;\dot{\rho},\rho^*,\tau^*))(1-F_a(t_{k-1};\dot{\rho},\rho^*,\tau^*))}\\
=&(1-G(t_k))(\frac{1}{1-F_a(t_k;\dot{\rho},\rho^*,\tau^*)}-\frac{1}{1-F_a(t_{k-1};\dot{\rho},\rho^*,\tau^*)})\\*
&\qquad {}\times [(Q_{a,X}(\dot{\rho};\rho^*,\tau^*))^\top d_X]_k.
\end{align*}
Then, we have
\begin{align*}
&\mathbb{E}[Z_a(\dot{\rho};\rho^*,\tau^*,\dot\eta)|A=a,X]\\
=&[\Sigma_{a,X}(\dot\rho;\rho^*,{\tau}^*,\dot{\eta})]^{-1}Q_{a,X}(\dot\rho;\rho^*,{\tau}^*)\mathbb{E}[R(\dot\rho;\rho^*,{\tau}^*)|A=a,X]\\
=&[\Sigma_{a,X}(\dot\rho;\rho^*,{\tau}^*,\dot{\eta})]^{-1} \Sigma_{a,X}(\dot\rho;\rho^*,{\tau}^*,\dot{\eta})d_X\\
=&d_X.
\end{align*}
Thus,
\begin{align*}
	&\mathbb{E} \left[ \frac{A}{e(X;\dot\gamma)}Z_1(\dot\rho;\rho^*,\tau^*,\dot\eta) - \frac{1-A}{1-e(X;\dot\gamma)}Z_0(\dot\rho;\rho^*,\tau^*,\dot\eta) \Big| X  \right]\\
	=& \frac{e(X)}{e(X;\dot\gamma)} \mathbb{E}[Z_1(\dot{\rho};\rho^*,\tau^*,\dot\eta)|A=1,X] - \frac{1-e(X)}{1-e(X;\dot\gamma)} \mathbb{E}[Z_0(\dot{\rho};\rho^*,\tau^*,\dot\eta)|A=0,X]\\
	=&  \frac{e(X)}{e(X;\dot\gamma)} d_X - \frac{1-e(X)}{1-e(X;\dot\gamma)} d_X=0.
\end{align*}

Therefore, under the union model $\widetilde{\mathcal{M}}_{\union}=\mathcal{M}_{\cir}\cap \{\widetilde{\mathcal{M}}_{\sop}\cup (\mathcal{M}_{\ps} \cap \mathcal{M}_{\cen})\}$, we have
\begin{align*}
&\mathbb{E}[U_{\cir}(\dot\rho;{\rho}^*, {\tau}^*, {\gamma}^*, {\eta}^*; \mathcal{W})]=\mathbb{E}[  \mathbb{E}\{U_{\cir}(\dot\rho;{\rho}^*, {\tau}^*, {\gamma}^*, {\eta}^*; \mathcal{W})|X\}]\\
=&\mathbb{E}\left[\mathcal{W}(X) \mathbb{E} \left\{ \frac{A}{e(X;\gamma^*)}Z_1(\dot\rho;\rho^*,\tau^*,\eta^*) - \frac{1-A}{1-e(X;\gamma^*)}Z_0(\dot\rho;\rho^*,\tau^*,\eta^*) \Big| X  \right\}\right]=0.
\end{align*}

Secondly, by applying Theorem 5.21 of \citet{van2000asymptotic}, we obtain the asymptotic normality of $\widehat{\rho}_{\DR}(\mathcal{W})$ and the corresponding influence function is
\begin{align}\label{supp-eq:if-cir}
	-\mathbb{E}\Big [\frac{\partial U_{\cir}(\rho;\rho^*,\tau^*,\gamma^*,\eta^*;\mathcal{W})}{\partial \rho} \Big |_{\rho=\dot{\rho}}\Big]^{-1} \widetilde{U}_{\cir}(\dot{\rho};\rho^*,\tau^*,\gamma^*,\eta^*;\mathcal{W}),
\end{align}
where
\begin{align*}
	&\widetilde{U}_{\cir}(\dot{\rho};\rho^*,\tau^*,\gamma^*,\eta^*;\mathcal{W})
	={U}_{\cir}(\dot{\rho};\rho^*,\tau^*,\gamma^*,\eta^*;\mathcal{W})\\
	&- \mathbb{E}\Big [\frac{\partial U_{\cir}(\dot\rho;\rho,\tau,\gamma^*,\eta^*;\mathcal{W})}{\partial (\rho^\top, \tau^\top)^\top}\Big |_{
		\begin{subarray}{c}
			\rho = \rho^* \\
			\tau=\tau^*
	\end{subarray}} \Big ] \Big \{  \mathbb{E}\Big [\frac{\partial S_{\cir}(\rho,\tau)}{\partial (\rho^\top, \tau^\top)^\top} \Big |_{\begin{subarray}{c}
			\rho = \rho^* \\
			\tau=\tau^*
	\end{subarray}}\Big ]\Big\}^{-1} S_{\cir}(\rho^*,\tau^*)\\
	&-\mathbb{E}\Big [\frac{\partial U_{\cir}(\dot\rho;\rho^*,\tau^*,\gamma,\eta^*;\mathcal{W})}{\partial \gamma}\Big |_{\gamma=\gamma^*} \Big] \Big \{ \mathbb{E}\Big[ -X\frac{\partial e(X;\gamma)}{\partial \gamma} \Big |_{\gamma=\gamma^*} \Big] \Big \}^{-1} X \{A-e(X;\gamma^*)\}\\
	&-\mathbb{E}\Big [\frac{\partial U_{\cir}(\dot\rho;\rho^*,\tau^*,\gamma^*,\eta;\mathcal{W})}{\partial \eta}\Big |_{\eta=\eta^*} \Big] \Big \{ \mathbb{E}\Big[ \frac{\partial S_{\cen}(\eta)}{\partial \eta} \Big |_{\eta=\eta^*} \Big] \Big \}^{-1} S_{\cen}(\eta^*).
\end{align*}

Finally, if $\widetilde{\mathcal{M}}_{\sop}$,
$\mathcal{M}_{\ps}$, and $\mathcal{M}_{\cen}$ are correctly
specified, it holds true that $(\rho^*,\tau^*,\gamma^*,\eta^*)=(\dot\rho,\dot\tau,\dot\gamma, \dot\eta)$. Then, we have
\begin{align*}
&\mathbb{E}\Big [\frac{\partial U_{\cir}(\dot\rho;\rho,\tau,\gamma^*,\eta^*;\widetilde{\mathcal W}_{\eff})}{\partial (\rho^\top, \tau^\top)^\top}\Big |_{
	\begin{subarray}{c}
		\rho = \rho^* \\
		\tau=\tau^*
\end{subarray}} \Big ]\\*
={}&\mathbb{E}\Big [\frac{\partial U_{\cir}(\dot\rho;\rho,\tau,\dot\gamma,\dot\eta;\widetilde{\mathcal W}_{\eff})}{\partial (\rho^\top, \tau^\top)^\top}\Big |_{
\begin{subarray}{c}
\rho = \dot\rho \\
\tau=\dot\tau
\end{subarray}} \Big ]\\
=&\mathbb{E}\Big [ \widetilde{\mathcal W}_{\eff}(X;\dot\rho,\dot\tau,\dot\gamma,\dot\eta)\Big\{\frac{A}{e(X;\dot\gamma)}\frac{\partial Z_1(\dot\rho;\rho,\tau,\dot\eta)}{\partial (\rho^\top, \tau^\top)^\top}\\*
&\qquad {}- \frac{1-A}{1-e(X;\dot\gamma)}\frac{\partial Z_0(\dot\rho;\rho,\tau,\dot\eta)}{\partial (\rho^\top, \tau^\top)^\top}\Big\}\Big |_{
	\begin{subarray}{c}
		\rho = \dot\rho \\
		\tau=\dot\tau
\end{subarray}}\\
&+\left\{\left(\frac{A}{e(X;\dot\gamma)}Z_1(\dot\rho;\dot\rho,\dot\tau,\dot\eta)-\frac{1-A}{1-e(X;\dot\gamma)}Z_0(\dot\rho;\dot\rho,\dot\tau,\dot\eta)\right)^\top \otimes \mathbf{I}_{p_\rho} \right\}\\*
&\qquad {}\times \frac{\partial \text{Vec}({ \widetilde{\mathcal W}_{\eff}(X;\rho,\tau,\dot\gamma,\dot\eta)})}{\partial (\rho^\top,\tau^\top)^\top} \Big |_{
	\begin{subarray}{c}
		\rho = \dot\rho \\
		\tau=\dot\tau
\end{subarray}} \Big ]\\
=&0,\\
&\mathbb{E}\Big [\frac{\partial U_{\cir}(\dot\rho;\rho^*,\tau^*,\gamma,\eta^*;\widetilde{\mathcal W}_{\eff})}{\partial \gamma}\Big |_{\gamma=\gamma^*} \Big]\\*
={}&\mathbb{E}\Big [\frac{\partial U_{\cir}(\dot\rho;\dot\rho,\dot\tau,\gamma,\dot\eta;\widetilde{\mathcal W}_{\eff})}{\partial \gamma}\Big |_{\gamma=\dot\gamma} \Big]\\
=&\mathbb{E}\Big [  \widetilde{\mathcal W}_{\eff}(X;\dot\rho,\dot\tau,\dot\gamma,\dot\eta)\Big\{ -\frac{A}{(e(X;\dot\gamma))^2}Z_1(\dot\rho;\dot\rho, \dot\tau,\dot\eta)\\*
&\qquad {}-\frac{1-A}{(1-e(X;\dot\gamma))^2}Z_0(\dot\rho; \dot\rho,\dot\tau,\dot\eta) \Big\}\frac{\partial e(X;\gamma)}{\partial \gamma}\Big |_{\gamma=\dot\gamma}\\
&+ \left\{\left(\frac{A}{e(X;\dot\gamma)}Z_1(\dot\rho;\dot\rho,\dot\tau,\dot\eta)-\frac{1-A}{1-e(X;\dot\gamma)}Z_0(\dot\rho;
\dot\rho,\dot\tau,\dot\eta)\right)^\top \otimes \mathbf{I}_{p_\rho} \right\}\\*
&\qquad {}\times \frac{\partial \text{Vec}({ \widetilde{\mathcal W}_{\eff}(X;\dot\rho,\dot\tau,\gamma,\dot\eta)})}{\partial \gamma} \Big |_{\gamma=\dot\gamma}\Big ]\\
=&0,\\
&\mathbb{E}\Big [\frac{\partial U_{\cir}(\dot\rho;\rho^*,\tau^*,\gamma^*,\eta;\widetilde{\mathcal W}_{\eff})}{\partial \eta}\Big |_{\eta=\eta^*} \Big]\\*
={}&\mathbb{E}\Big [\frac{\partial U_{\cir}(\dot\rho;\dot\rho,\dot\tau,\dot\gamma,\eta;\widetilde{\mathcal W}_{\eff})}{\partial \eta}\Big |_{\eta=\dot\eta} \Big]\\
=&\mathbb{E}\Big [ \widetilde{\mathcal W}_{\eff}(X;\dot\rho,\dot\tau,\dot\gamma,\dot\eta)\Big\{\frac{A}{e(X;\dot\gamma)}\frac{\partial Z_1(\dot\rho;\dot\rho,\dot\tau,\eta)}{\partial \eta}\\*
&\qquad {}- \frac{1-A}{1-e(X;\dot\gamma)}\frac{\partial Z_0(\dot\rho;\dot\rho,\dot\tau,\eta)}{\partial \eta}\Big\}\Big |_{\eta=\dot\eta}\\
&+ \left\{\left(\frac{A}{e(X;\dot\gamma)}Z_1(\dot\rho;\dot\rho,\dot\tau,\dot\eta)-\frac{1-A}{1-e(X;\dot\gamma)}Z_0(\dot\rho;\dot\rho,\dot\tau,\dot\eta)\right)^\top \otimes \mathbf{I}_{p_\rho} \right\}\\*
&\qquad {}\times \frac{\partial \text{Vec}({ \widetilde{\mathcal W}_{\eff}(X;\dot\rho,\dot\tau,\dot\gamma,\eta)})}{\partial \eta} \Big |_{\eta=\dot\eta}\Big ]\\
=&0.
\end{align*}
Therefore, the influence function of $\widehat{\rho}_{\DR}(\widetilde{\mathcal W}_{\eff})$ becomes $$-\mathbb{E}\Big [\frac{\partial U_{\cir}(\rho;\dot\rho,\dot\tau,\dot\gamma,\dot\eta;\widetilde{\mathcal{W}}_{\eff})}{\partial \rho} \Big |_{\rho=\dot{\rho}}\Big]^{-1} {U}_{\cir}(\dot{\rho};\dot\rho,\dot\tau,\dot\gamma,\dot\eta;\widetilde{\mathcal{W}}_{\eff}).$$

By an argument analogous to that used in the proof of Proposition \ref{prop:sur:dr}, with the additional orthogonality of the censoring tangent space, this is the efficient influence function under \(\widetilde{\mathcal M}_{\union}\). Therefore, if $\widetilde{\mathcal{M}}_{\sop}$,
$\mathcal{M}_{\ps}$, and $\mathcal{M}_{\cen}$ are correctly
specified, \(\widehat{\rho}_{\DR}(\widetilde{\mathcal W}_{\eff})\) attains the semiparametric efficiency bound under \(\widetilde{\mathcal M}_{\union}\).
\end{proof}

The covariance matrix of \(\widehat{\rho}_{\DR}(\mathcal W)\) is estimated using the influence function in \eqref{supp-eq:if-cir}, and the Wald confidence intervals are constructed accordingly.

\subsection{Semiparametric estimation under \texorpdfstring{$C\ind T \mid (A,X)$}{C independent of T given (A,X)}}\label{supp-sec:cir-censoring}
In this part, we assume $C\ind T \mid (A,X)$ and let $g_a(t;X)=\pr(C=t \mid X,A=a)$ and $G_a(t_k;X)=\sum_{i=1}^{k-1}g_a(t_i;X)$. Let \(\widetilde{\Sigma}^\dagger_{A,X}\) be the diagonal matrix with $k$-th diagonal element \[\widetilde{\sigma}^{k,\dagger}_{A,X}=\{1-G_A(t_k;X)\}
\left\{
\frac{1}{1-F_A(t_k;X)}
-
\frac{1}{1-F_A(t_{k-1};X)}
\right\},\]
and \(\Sigma^{\dagger}_{A,X}
=
Q_{A,X}
\widetilde{\Sigma}^{\dagger}_{A,X}
Q_{A,X}^{\top}.\) We also define
\begin{align*}
	&\overline{\Sigma}^{\dagger}_{X}
	=
	e(X)\Sigma^{\dagger}_{1,X}
	+
	\{1-e(X)\}\Sigma^{\dagger}_{0,X},\\
	&\Omega^{\dagger}_{X}
	=
	e(X)\{1-e(X)\}
	\Sigma^{\dagger}_{1,X}
	\big(\overline{\Sigma}^{\dagger}_{X}\big)^{-1}
	\Sigma^{\dagger}_{0,X},\\
	&\widetilde{\mathcal W}^{\dagger}_{\mathrm{eff}}(X)
	=
	\mathcal F_{\mathrm{cir},X}^{\top}
	\Omega^{\dagger}_{X},\\
	& Z^{\dagger}_a(\rho)
	=
	\big(\Sigma^{\dagger}_{a,X}\big)^{-1}
	J_{a,X}(\rho),
	\qquad a=0,1.
\end{align*}

\begin{proposition}\label{supp-prop:effScore-cir}
Under the observed-data law restricted by the model $\mathcal{M}_{\cir}$, the efficient score for $\rho$ is
\begin{align*}
\widetilde{S}^*(\rho)=\widetilde{\mathcal{W}}^\dagger_{\eff}(X)\left[\frac{A}{e(X)}Z^\dagger_1(\rho)-\frac{1-A}{1-e(X)}Z^\dagger_0(\rho)\right],
\end{align*}
where $\widetilde{\mathcal{W}}^\dagger_{\eff}(X)$ is the efficient $p_{\rho}\times m$ weighting matrix, and $Z^\dagger_a(\rho)$ is an arm-specific $m$-dimensional residual.
\end{proposition}

\begin{proof}[Proof of Proposition \ref{supp-prop:effScore-cir}]

The proof follows from the proof of Theorem \ref{thm:sur:effscore-CIR} after replacing \(G(t;X)\) by \(G_A(t;X)\) and \(\widetilde{\Sigma}_{A,X}\) by \(\widetilde{\Sigma}^\dagger_{A,X}\).

\end{proof}

Based on Proposition \ref{supp-prop:effScore-cir}, a corresponding doubly robust estimator can be constructed.

\section*{Supplementary Tables}
\renewcommand{\thetable}{S\arabic{table}}
\setcounter{table}{0}

\begin{table}[htbp!]
	\centering
	\caption{Monte Carlo results of the proposed MLE and DR estimators. The true values for $\alpha_0$ and $\alpha_1$ are 0 and $1$, respectively, for both examples. Sample size = 1000}
	\label{supp-tab:dr1000}
	\begin{threeparttable}
		\begin{tabular}{llrrrr}
			\toprule
			& & \multicolumn{2}{c}{Example 1} & \multicolumn{2}{c}{Example 2} \\
			\cmidrule(lr){3-4} \cmidrule(lr){5-6}
			& & \multicolumn{1}{c}{$\alpha_0$} & \multicolumn{1}{c}{$\alpha_1$}
			& \multicolumn{1}{c}{$\alpha_0$} & \multicolumn{1}{c}{$\alpha_1$} \\
			\midrule
			\multicolumn{6}{l}{\textbf{Bias (SE)}} \\
			& mle.bth & $-0.003\;(0.003)$ & $\phantom{-}0.006\;(0.003)$ & $\phantom{-}0.001\;(0.004)$ & $\phantom{-}0.001\;(0.004)$ \\
			& mle.bad & $\phantom{-}0.256\;(0.003)$ & $\phantom{-}0.076\;(0.003)$ & $\phantom{-}0.109\;(0.004)$ & $\phantom{-}0.112\;(0.004)$ \\
			& dr.bth  & $-0.003\;(0.003)$ & $\phantom{-}0.007\;(0.004)$ & $-0.005\;(0.005)$ & $\phantom{-}0.008\;(0.005)$ \\
			& dr.psc  & $-0.003\;(0.003)$ & $\phantom{-}0.008\;(0.004)$ & $-0.006\;(0.005)$ & $\phantom{-}0.010\;(0.005)$ \\
			& dr.sop  & $-0.002\;(0.003)$ & $\phantom{-}0.007\;(0.004)$ & $-0.003\;(0.005)$ & $\phantom{-}0.008\;(0.005)$ \\
			& dr.bad  & $-0.100\;(0.003)$ & $-0.025\;(0.004)$ & $-0.183\;(0.005)$ & $\phantom{-}0.035\;(0.005)$ \\
			\addlinespace
			\multicolumn{6}{l}{\textbf{SD Accuracy\textsuperscript{*}}} \\
			& mle.bth & 0.986 & 0.974 & 0.993 & 1.035 \\
			& mle.bad & 0.971 & 0.984 & 1.006 & 1.017 \\
			& dr.bth  & 0.986 & 0.978 & 0.978 & 1.028 \\
			& dr.psc  & 0.973 & 0.974 & 0.983 & 1.024 \\
			& dr.sop  & 0.983 & 0.969 & 0.981 & 1.034 \\
			& dr.bad  & 0.972 & 0.971 & 0.911 & 0.995 \\
			\addlinespace
			\multicolumn{6}{l}{\textbf{Coverage\textsuperscript{**}}} \\
			& mle.bth & 0.944 & 0.938 & 0.944 & 0.968 \\
			& mle.bad & 0.036 & 0.846 & 0.748 & 0.778 \\
			& dr.bth  & 0.944 & 0.946 & 0.934 & 0.950 \\
			& dr.psc  & 0.950 & 0.938 & 0.934 & 0.956 \\
			& dr.sop  & 0.952 & 0.946 & 0.930 & 0.948 \\
			& dr.bad  & 0.740 & 0.922 & 0.580 & 0.944 \\
			\bottomrule
		\end{tabular}
		\begin{tablenotes}
			\item[*] SD Accuracy = Average estimated SD / Monte Carlo SD.
			\item[**] Confidence level = 95\%.
		\end{tablenotes}
	\end{threeparttable}
\end{table}

\begin{table}[htbp!]
	\centering
	\caption{Monte Carlo results of the proposed MLE and DR estimators.
		The true values for $\rho_0$ and $\rho_1$ are 0.8 and 1.4, respectively,
		for both examples. Sample size = 1000}
	\label{supp-tab:cir_1000}
	\begin{threeparttable}
		\begin{tabular}{llrrrr}
			\toprule
			& & \multicolumn{2}{c}{Example 1} & \multicolumn{2}{c}{Example 2} \\
			\cmidrule(lr){3-4} \cmidrule(lr){5-6}
			& & \multicolumn{1}{c}{$\rho_0$} & \multicolumn{1}{c}{$\rho_1$}
			& \multicolumn{1}{c}{$\rho_0$} & \multicolumn{1}{c}{$\rho_1$} \\
			\midrule
			\multicolumn{6}{l}{\textbf{Bias (SE)}} \\
			& mle.all & $-0.014\;(0.021)$ & $\phantom{-}0.019\;(0.018)$ & $-0.015\;(0.021)$ & $\phantom{-}0.009\;(0.017)$ \\
			& mle.bad & $-0.107\;(0.019)$ & $\phantom{-}1.148\;(0.018)$ & $-0.016\;(0.019)$ & $\phantom{-}1.166\;(0.018)$ \\
			& dr.all  & $\phantom{-}0.042\;(0.022)$ & $-0.027\;(0.018)$ & $-0.007\;(0.021)$ & $\phantom{-}0.006\;(0.018)$ \\
			& dr.psc  & $-0.003\;(0.023)$ & $\phantom{-}0.017\;(0.020)$ & $-0.007\;(0.023)$ & $\phantom{-}0.001\;(0.020)$ \\
			& dr.sop  & $\phantom{-}0.035\;(0.022)$ & $-0.032\;(0.018)$ & $-0.003\;(0.022)$ & $\phantom{-}0.000\;(0.018)$ \\
			& dr.bad  & $-0.995\;(0.021)$ & $\phantom{-}0.532\;(0.019)$ & $-1.031\;(0.021)$ & $\phantom{-}0.521\;(0.019)$ \\
			\addlinespace
			\multicolumn{6}{l}{\textbf{SD Accuracy\textsuperscript{*}}} \\
			& mle.all & 1.004 & 0.972 & 1.008 & 0.976 \\
			& mle.bad & 0.985 & 0.953 & 1.014 & 1.000 \\
			& dr.all  & 1.001 & 1.005 & 1.006 & 0.968 \\
			& dr.psc  & 1.035 & 1.011 & 1.003 & 0.961 \\
			& dr.sop  & 1.002 & 1.007 & 0.999 & 0.979 \\
			& dr.bad  & 0.989 & 0.981 & 0.994 & 0.964 \\
			\addlinespace
			\multicolumn{6}{l}{\textbf{Coverage\textsuperscript{**}}} \\
			& mle.all & 0.972 & 0.940 & 0.954 & 0.942 \\
			& mle.bad & 0.946 & 0.148 & 0.954 & 0.150 \\
			& dr.all  & 0.956 & 0.944 & 0.956 & 0.948 \\
			& dr.psc  & 0.968 & 0.956 & 0.968 & 0.938 \\
			& dr.sop  & 0.950 & 0.946 & 0.954 & 0.946 \\
			& dr.bad  & 0.428 & 0.762 & 0.402 & 0.750 \\
			\bottomrule
		\end{tabular}
		\begin{tablenotes}
			\item[*] SD Accuracy = Average estimated SD / Monte Carlo SD.
			\item[**] Confidence level = 95\%.
		\end{tablenotes}
	\end{threeparttable}
\end{table}

\begin{table}[htbp!]
	\centering
	\caption{Estimates of $\rho_1$ and $\rho_2$ in the selected cohort from the HIP Breast Cancer Screening Trial using maximum likelihood and doubly robust methods}
	\label{supp-tab:realdata-cir}
	\small
	\setlength{\tabcolsep}{1.5pt}
	\begin{tabular}{@{}lcrrrr@{}}
		\toprule
		\multirow{2}{*}{Contrast} & \multirow{2}{*}{Parameter}
		& \multicolumn{2}{c}{MLE} & \multicolumn{2}{c}{DR} \\
		\cmidrule(lr){3-4}\cmidrule(lr){5-6}
		& & \multicolumn{1}{c}{Estimate (95\% CI)} & \multicolumn{1}{c}{$P$-value}
		& \multicolumn{1}{c}{Estimate (95\% CI)} & \multicolumn{1}{c}{$P$-value} \\
		\midrule
		\multirow{2}{*}{\shortstack[c]{Control vs \\ Screening invitation}} & $\rho_1$ & $-0.985\;(-2.680,\phantom{-}0.711)$ & $0.255$ & $-0.961\;(-2.538,\phantom{-}0.617)$ & $0.233$ \\
		& $\rho_2$ & $\phantom{-}1.150\;(\phantom{-}0.300,\phantom{-}2.000)$ & $0.008$ & $\phantom{-}1.401\;(\phantom{-}0.463,\phantom{-}2.340)$ & $0.003$ \\
		\bottomrule
	\end{tabular}
\end{table}
\FloatBarrier

\end{document}